**Heterogeneous evolution of the galaxy and the origin of the short-lived nuclides in the early solar system.** T. Kaur and S. Sahijpal, Department of Physics, Panjab University, Chandigarh, India 160014.

**Abstract:**

We present galactic chemical evolution (GCE) models of the short-lived radionuclides (SLRs), $^{26}$Al, $^{36}$Cl, $^{41}$Ca, $^{53}$Mn and $^{60}$Fe, across the entire Milky Way galaxy. The objective is to understand the spatial and temporal distribution of the SLRs in the galaxy. The gamma-ray observations infer widespread distribution of $^{26}$Al and $^{60}$Fe across the galaxy. The signatures of the SLRs in the early solar system (ESS) are found in meteorites. We present homogeneous GCE simulation models for SLRs across the galaxy. We also develop a set of heterogeneous GCE models to understand the evolution of the galaxy within independent spatial grids of area, 0.1-1 kpc$^2$. These grids evolve distinctly in terms of nucleosynthetic contributions of massive stars. We succeeded in simulating the formation and evolution of generations of stellar clusters/association. Based on the formulation, we provide a novel method to amalgamate the origin of the solar system with the gradual evolution of the galaxy along with a self-consistent origin of SLRs. We explore the possibility of the birth of the solar system in an environment where one of the stellar clusters formed ≥25 Million years earlier. The decaying $^{53}$Mn and $^{60}$Fe remnants from the evolved massive stars from the cluster probably contaminated the local medium associated with the presolar molecular cloud. A Wolf-Rayet wind from a distant massive star, belonging to a distinct cluster, probably contributed, $^{26}$Al (and $^{41}$Ca) to the presolar cloud. The irradiation production of $^{7,10}$Be and $^{36}$Cl occurred later in ESS.

# 1. Introduction

The formation and evolution of galaxies initiated within 1 Gyr (Giga years) from the origin of the universe approximately 13.7 Gyr ago. Various generations of stars are formed inside the galaxies and evolved over the galactic time-scales. The primordial gas that evolved subsequent to the primordial nucleosynthesis was mostly Hydrogen and Helium rich with approximate mass fractions of 0.75 and 0.25, respectively. However, the spectroscopic observations of stars (including the sun) and interstellar medium reveal a widespread distribution of elements heavier than Lithium. The stellar nucleosynthesis within several generations of stars formed and evolved over the galactic time-scales resulted in the gradual increase in the elemental abundance distribution among the galaxies. The galactic chemical evolution (GCE) models deal with the understanding of the formation and the evolution of stars along with their nucleosynthetic contribution to the interstellar medium (see e.g., Pagel 1997; Matteucci 2003). The GCE models incorporate the dynamical aspects related with the formation and evolution of the galaxy along with the homogenization processes operating within the galaxy on account of interstellar gas mixing and stellar migration (Matteucci & François 1989; Chiappini, Matteucci & Romano 2001; Kobayashi et al. 2006; Minchev et al. 2012, 2015; Sahijpal & Gupta 2013; Grisoni, Spitoni & Matteucci 2018; Sahijpal & Kaur 2018).

During the gradual evolution of the Milky Way galaxy, the formation of the solar system commenced around 4.56 billion years ago by the gravitational collapse of a presolar molecular cloud. A wide range of astrophysical scenarios have been proposed for the origin of the solar system (see e.g., Boss 1995; Cameron et al. 1995; Wasserburg et al. 2006; Sahijpal & Soni 2006; Gaidos et al. 2009; Gounelle et al. 2009; Huss et al. 2009; Sahijpal and Gupta 2009; Dauphas and Chaussidon 2011; Boss 2017; Liu 2017; Lugaro, Ott & Keresztüri 2018; Jones et al. 2019). Evolved stars ranging from massive Wolf-Rayet (WR) stars, Supernovae (SN Ia, Ib/c, SN II) and Asymptotic Giant Branch (AGB) stars have been proposed along with distinct stellar environments associated with the birth of the solar system. The definite association of evolved star(s) with the formation of the solar system is necessitated by the presence of several short-lived nuclides (SLRs), e.g., $^{26}$Al ($\tau$, mean life ~1.05 Myr), $^{36}$Cl ($\tau$ ~0.43 Myr), $^{41}$Ca ($\tau$ ~0.15 Myr), $^{53}$Mn ($\tau$ ~5.34 Myr) and $^{60}$Fe ($\tau$ ~3.75 Myr) in the early solar system (ESS) (Lee, Papanastassiou & Wasserburg 1976, 1977; Clayton, Hartmann & Leising 1993; MacPherson, Davis & Zinner 1995; Russell, Gounelle & Hutchison 2001; Jacobsen et al. 2008; Krot et al. 2008, 2012; Amelin et al. 2010; Bouvier & Wadhwa 2010; Connelly et al. 2012; Tang & Dauphas 2012, 2015; Mishra & Goswami 2014; Telus et al. 2016; Liu 2017; Tang et al. 2017; Tissot, Dauphasa & Grove 2017; Trappitsch et al. 2018).

The presence of the short-lived radionuclides in the early solar system (ESS) has been deduced on the basis of enrichment in the abundance of their daughter nuclides over the standardly accepted value in various meteoritic phases. Some of these SLRs have been extensively used as chronometers for understanding the ESS processes ranging from the condensation of the earliest solar system grains to the formation and planetary-scale differentiation of planetesimals and other planetary bodies. $^{26}$Al, in particular, has been considered as a potent heat source for a wide range of thermal processes operating within the planetesimals and other planetary bodies in the ESS (Urey 1955; Grimm & McSween 1993; Merk, Breuer & Spohn 2002; Sahijpal, Soni & Gupta 2007; Neumann, Breuer & Spohn 2014; Bhatia & Sahijpal 2016). Based upon the analyses of Al-Mg isotopic systematics in large numbers of Calcium–Aluminum-rich Inclusions (CAIs), the earliest formed solar system grains, a canonical value of ~5×10$^{-5}$ for $^{26}$Al/$^{27}$Al was estimated in the ESS (MacPherson, Davis & Zinner 1995). Due to the subsequent extensive high-precision studies of Al-Mg isotopic systematics by several groups (Russell et al. 1996; Galy, Hutcheon & Grossman 2004; Bizzarro, Baker & Haack 2005; Young et al. 2005; Thrane, Bizzarro & Baker 2006), the canonical value of $^{26}$Al/$^{27}$Al has been eventually revised to a well-accepted value of (5.23±0.13)×10$^{-5}$ (Jacobsen et al. 2008).

Apart from $^{26}$Al, one of the other most crucial SLR studied in the ESS is $^{60}$Fe. The initially proposed high estimates of $^{60}$Fe in chondrules, with an initial $^{60}$Fe/$^{56}$Fe value of ~1×10$^{-6}$ (Mostefaoui, Lugmair & Hoppe 2005) indicated a possible role of $^{60}$Fe as a potent heat source for thermal processing of planetesimals. However, the revised initial $^{60}$Fe/$^{56}$Fe ratio of ~1.01×10$^{-8}$ (Tang & Dauphas 2012, 2015) minuscule any possible role of $^{60}$Fe as a heat source (see e.g., Bhatia & Sahijpal 2016). The *in situ* measurements from unequilibrated ordinary chondrites (UOCs) by various groups provide a diverse range of $^{60}$Fe/$^{56}$Fe ratios. This include the initial ratios of ~7×10$^{-7}$ (Mishra & Chaussidon 2014), (2-8)×10$^{-7}$ (Mishra & Goswami 2014), (8-11)×10$^{-7}$ (Mishra, Marhas & Sameer 2016) and 5×10$^{-8}$ to 3×10$^{-7}$ (Telus et al. 2012, 2018). However, the recent work by Trappitsch et al. (2018) using *Resonance Ionization Mass Spectrometry* (RIMS) measurements provide a low initial value of (6.4±11.9)×10$^{-8}$ for $^{60}$Fe/$^{56}$Fe at the time of formation of CAIs. We confine our present discussion to the initial $^{60}$Fe/$^{56}$Fe ratio of ~1.01×10$^{-8}$ (Tang & Dauphas 2012, 2015) for the ESS.

Several estimates are available for the initial abundance of $^{53}$Mn in the ESS phases. The inferred initial $^{53}$Mn/$^{55}$Mn values by various groups are in general agreement with each other. The recommended value of (7±1)×10$^{-6}$ by Tissot et al. (2017) will be considered as a reference in the present work. Tissot et al. (2017) made a comparison of the available values of $^{53}$Mn/$^{55}$Mn in literature, with the values ranging from (5.4±2.4)×10$^{-6}$ to (8.5±1.5)×10$^{-6}$ (Lugmair & Galer 1992; Göpel, Manhes & Allegre, 1994; Lugmair & Shukolyukov 1998; Glavin et al. 2004; Amelin 2008; Brennecka & Wadhwa 2012; Connelly et al. 2012). The initial estimates of $^{36}$Cl vary widely in literature. The reference value of (2.44 ± 0.65) ×10$^{-5}$ for the initial $^{36}$Cl/$^{35}$Cl, considered in the present work, is derived from Tang et al. (2017) on the basis of the analysis of the Curious Marie CAI (Lugaro et al. 2018). The initial estimates of $^{41}$Ca has been revised recently with the use of high precision *Secondary Ion Mass Spectrometer* (SIMS) (Liu et al. 2012). The measurements performed on EGG3 pyroxene, for example, on a large geometry SIMS-ims1270 by Ito, Nagasawa & Yurimoto (2006) yielded initial $^{41}$Ca/$^{40}$Ca value of ~4.1×10$^{-9}$ compared to a small geometry SIMS-ims4f value of ~1.2×10$^{-8}$ for $^{41}$Ca/$^{40}$Ca by Sahijpal, Goswami & Davis (2000) on the same sample. In the present work, we adopt an initial value of (4.6±1.9)×10$^{−9}$ for $^{41}$Ca/$^{40}$Ca (Liu 2017) as a reference value.

The origin of SLRs remains a debatable issue despite several viable proposals that include stellar nucleosynthesis, galactic chemical evolution, and irradiation scenarios. The time scales involved in the mixing of the interstellar medium (ISM) are much longer than the mean life of these SLRs. This limits a substantial stellar nucleosynthetic contribution of SLRs to the ESS from the gradual evolution of the galaxy. The observed abundances of SLRs, especially, $^{26}$Al in the ESS require a prompt stellar nucleosynthetic contribution. This becomes even more crucial as it has been demonstrated that irradiation scenario, especially, the proposed X-wind irradiation scenario cannot reproduce the observed canonical value of $^{26}$Al/$^{27}$Al unless superflares with X-ray luminosities ≥10$^{33-34}$ erg s$^{-1}$ are invoked for complete vaporization and homogenization of irradiated protoCAIs over a large scale by the irradiation environment prevailing in the vicinity of the early active sun (Sahijpal & Gupta 2009). The irradiation environment prevailing in the ESS can nonetheless explain the irradiation origin of $^{7,10}$Be, $^{41}$Ca, $^{36}$Cl and $^{53}$Mn (Leya, Halliday & Wieler 2003; Gounelle et al. 2006; Sahijpal & Gupta 2009).

Various stellar nucleosynthetic processes can contribute towards the predicted abundance of the SLRs (Cameron et al. 1995; Wasserburg et al. 2006; Sahijpal & Soni 2006; Gaidos et al. 2009; Gounelle et al. 2009; Huss et al. 2009; Sahijpal and Gupta 2009; Dauphas and Chaussidon 2011; Boss 2017; Bojazi & Meyer 2017; Dwarkadas et al. 2017; Liu 2017; Lugaro et al. 2018; Jones et al. 2019). The deduced stellar nucleosynthetic yields and the inferred initial abundances of SLRs in various meteorites impose constraints on the theoretical formulation regarding the formation of the solar system. The majority of the stellar nucleosynthetic hypotheses argue in favor of the injection of SLRs from a stellar source into the presolar molecular cloud prior to the formation of the solar system. The possibility of AGB (Asymptotic Giant Branch) star (3-8 M$_\odot$) as a source of SLRs, $^{26}$Al, $^{107}$Pd, and $^{182}$Hf, present in the ESS, has been explored by Wasserburg et al. (2006). However, it has been argued earlier that AGB stars are very rarely found near star-forming regions (Kastner & Myers 1994), and hence, there are few chances of the contribution of SLRs to the ESS from AGB stars. Massive stars synthesize substantial amounts of $^{26}$Al, $^{41}$Ca, $^{36}$Cl, $^{53}$Mn and $^{60}$Fe. Several groups have considered the contributions of SLRs from massive stars to the ESS (Cameron et al. 1995; Sahijpal & Soni 2006;

Gaidos et al. 2009; Gounelle et al. 2009; Huss et al. 2009; Sahijpal & Gupta 2009; Dwarkadas et al. 2017; Liu 2017; Lugaro et al. 2018; Jones et al. 2019). Massive stars ($\geq 33 M_\odot$) evolve through the Wolf-Rayet (WR) phase that can contribute a substantial amount of $^{26}$Al. Based on the initial hypothesis proposed by Cameron et al. (1995), Sahijpal & Soni (2006) discussed in detail the nucleosynthetic contribution by a WR star that could have been either a single star or a star in a binary system. They considered the possibilities of Wolf-Rayet wind with both post-core-collapse supernova and without the supernova. The later models do not produce $^{60}$Fe that is essentially produced during a core-collapse supernova. It was demonstrated that the canonical abundance of $^{26}$Al/$^{27}$Al in the ESS can be obtained from the WR winds without producing any unobserved stable isotopic anomaly among CAIs (Sahijpal & Soni 2006).

Young (2014, 2016) also suggested that the SLR abundance in the ESS can be explained by the contribution from WR winds without any contribution from the supernova injection. The uncertainty in the initial $^{60}$Fe/$^{56}$Fe in the ESS kept the role of a core-collapse supernova contribution of $^{60}$Fe uncertain for a long time. However, the recent evidence for the low value of initial $^{60}$Fe/$^{56}$Fe (Trappitschet al. 2018) rejects the hypothesis involving a core-collapse supernova contribution of $^{60}$Fe to the ESS (Jones et al. 2019). Based on the high $^{26}$Al/$^{27}$Al and a low value of $^{60}$Fe/$^{56}$Fe observed in the solar system, Dawakadas et al. (2017) developed a formulation in which the formation of solar system occurred at the edge of a WR bubble. The evolving massive star produced $^{26}$Al which survived and entered the ESS. The $^{60}$Fe produced by SN could not reach the solar system leaving $^{60}$Fe/$^{56}$Fe around the galactic value.

An alternative hypothesis dealing with the injection of SLRs in an already existing protoplanetary disc by an evolved star(s) has also been proposed (Hester, Healy & Desch 2004; Ouellette, Desch & Hester 2007). However, in the wake of the recent low estimates of $^{60}$Fe in the ESS, even the alternative scenarios cannot invoke major contributions of a core-collapse supernova. Nonetheless, in general, a spiked stellar contribution of SLRs, specially, $^{26}$Al, to the ESS rather than a gradual contribution from galactic chemical evolution (Huss et al. 2009; Sahijpal 2014a; Fujimoto, Krumholz & Tachibana 2018; Lugaro et al. 2018; Jones et al. 2019) is considered to have resulted in the observed abundances of at least some of the SLRs. However, the origin of the solar system has to be understood in terms of the successive formation and evolution of stellar cluster(s) during the gradual evolution of the Milky Way galaxy (see e.g., Sahijpal 2014a). In this context, the recent chemodynamical evolutionary model provides an account of the heterogeneous chemical evolution of SLRs in the galaxy (Fujimoto et al. 2018). In this study, the nucleosynthesis ejecta due to the ongoing evolution of the galaxy is considered to have enriched the interstellar medium (ISM) with SLRs. The giant molecular clouds are formed from ISM. The gravitational collapse of dense molecular cloud cores results in the formation of stars with entrapped SLRs. The large molecular clouds can host a wide number of active stars forming regions where the stars are formed within stellar clusters. An attempt is made in the present work to quantify the production rates of the SLRs across the entire Milky Way galaxy in order to understand the origin of the observed galactic distribution of SLR (Diehl et al. 2006a, b) and the contribution of SLRs to the planetary systems, e.g., the solar system, formed across the galaxy at distinct times. Further, we make an attempt to understand the formation of the solar system in context with the ongoing evolution of the galaxy. In general, our major objective is to reconcile the origin of the solar system with the gradual heterogeneous evolution of the galaxy.

As majority of the stars are formed within stellar clusters, the time period of star formation greatly depends upon the life cycle of the giant molecular cloud complexes (Lada & Lada 2003; Smith 2006; Weidner & Kroupa 2006; Smith & Brooks 2007; Feigelson & Townsley 2008; Kroupa 2008; Lada, Lombardi & Alves 2010; Clark et al. 2012; Howard, Pudritz & Harris 2014). There is a good understanding regarding star formation and stellar feedback processes but the formation of large molecular cloud complexes is still not very clear. These cloud complexes host several active sites for star formation during their typical lifetime of tens of million years, with mass up to $10^6\,M_\odot$ and spatial dimensions of several tens of parsecs. Due to gravitational instabilities and turbulence inside these cloud complexes, the clouds get fragmented into clumps of high density. The gravitational collapse of these clumps results in the formation of stars. The efficiency of the process of conversion of gas of the molecular cloud into the stars depends upon the prevailing turbulence, the magnetic field, and stellar feedback processes.

Since the stars are mostly formed inside stellar clusters so there is a definite probability that the formation of the solar system also occurred inside a stellar cluster along with stars of different masses (Smith 2006; Weidner & Kroupa 2006; Smith & Brooks 2007). The observational evidence of the eccentric orbits of trans-Neptunian objects could have resulted from the early encounters of the Sun with other companion stars from the cluster. (Thies, Kroupa & Thies 2005; Trujillo & Sheppard 2014; Pfalzner et al. 2015). The stellar mass distribution within a stellar cluster follows an initial mass function, IMF (Lada & Lada 2003). After the commencement of the formation of a cluster, the massive stars (e.g, a star with mass > 60 $M_\odot$) evolve fast over a time-scale of 3-5 Myr compared to the contemporary stars with low and intermediate masses. The massive stars can eventually disrupt further star formation in the cluster either through their supernova shock waves or high wind losses and ultraviolet radiation fields. On the other hand, the shock waves from the supernova or stellar winds associated with Wolf-Rayet stars can initiate the gravitational collapse of the surrounding molecular clouds. The dynamics governed by distance and the density of the star-forming region determines the ongoing episode(s) of star formation in the vicinity of massive stars. This might also have played a key role in the birth of the solar system. The stellar winds from the massive stars influence the surrounding and can inject newly synthesized matter into the accretion disc of the proto-stars (Hester et al. 2004; Ouellette et al. 2007).

In order to disentangle the obscurity of the origin of the solar system, in the present work, we have performed GCE simulations of the entire Milky Way galaxy in which the solar system originates as a natural consequence of the galactic evolution around 4.56 Gyr ago. The GCE model deals with the time evolution of the elemental and isotopic abundance by considering the accretion history of the galaxy, the star formation rate and the stellar nucleosynthetic inventories from various generations of stars formed over the galactic time-scale (Matteucci & François 1989; Chiappini et al. 2001; Kobayashi et al. 2006; Minchev et al. 2012, 2015; Sahijpal & Gupta 2013; Grisoni et al. 2018; Sahijpal & Kaur 2018). The conventional GCE models deal with solving the complex integrodifferential equations for the stable isotopes. In an alternative approach (Sahijpal & Gupta 2013; Sahijpal & Kaur 2018), the galaxy is evolved as an ensemble of numerous generations of stars. Sahijpal & Kaur (2018) recently developed the GCE model based on N-body Monte Carlo numerical simulations across the 2-18 kpc radial extent of the galaxy. Numerous generations of stars are formed according to the stellar birth rate function, and these stars are evolved over mass-dependent life-spans. The stars synthesize various elements at different stages during their life and enrich the interstellar medium (ISM) gas which further produces the next generation of stars. Over the galactic evolution, the formation of the solar system took place ~4.56 Gyr ago at an assumed distance of ~8.5 kpc from the galactic center. One of the major objectives of the GCE models is to reproduce the stellar elemental abundance distribution (Sahijpal & Kaur 2018) across the galaxy. In the present work, we have used the GCE formulation recently developed by Sahijpal & Kaur (2018) to understand the gradual abundance evolution of SLRs based on homogeneous as well as heterogeneous GCE models.

During the evolution of the galaxy at instants when the stellar inventories are ejected into the ISM, the mixing length and time-scales are very important to understand the dynamics related to the galactic evolution of the stable and SLR inventories. In the majority of the earlier GCE models, the instantaneous mixing approximation was considered which allows the nucleosynthetic inventories to mix within the entire defined annular ring instantaneously after the lifetime evolution of stars. However, the observational evidence suggest that it will take significant time for the stellar ejecta to mix with the ISM. In order to deal with the complications related to the mixing time scales, Sahijpal (2013) presented an inhomogeneous GCE model for stable nuclides in the solar neighborhood. In this work, the solar annular ring at a distance of 7-8.8 kpc from the galactic center is divided into the spatial grids of size 1-2 kpc$^2$ each. The stars inside each grid were formed and evolved in isolation from the other grids. The special feature of this inhomogeneous GCE approach is that the nucleosynthetic inventories contribute to the respective grid alone where the birth of star took place. This assumption specifically holds for massive stars that evolve rapidly and explode as supernovae (SN II and SN Ib/c). In this manner, the localized mixing is incorporated to partially reproduce the observed elemental heterogeneities across the galaxy. The mixing time-scales vary over the spatial dimensions, average temperature, and composition of the ISM. The proposed hypothesis suggests that inside a disc galaxy the dynamical processes along with the turbulent flows induced by the differential rotation of the galaxy should remove the azimuthal heterogeneities over the galactic spatial scales of 1 - 10 kpc within 1 Gyr (Roy & Kunth 1994). The analysis of various hydrodynamic processes at a scale of 100 pc to 1000 pc shows that the collision of clouds and the expanding shells of the massive stars associations along with differential rotation and triggered star formation homogenize the matter in $10^8$ years. On the smaller spatial scales of 1 pc to 100 pc, the mixing due to turbulent diffusion can homogenize the cold clouds within a couple of million years. The supernova driven winds remove the heterogeneities at a scale of 500 pc over a few 100 Myr (de Avillez & Mac Low 2002). The associated findings suggest that the larger the spatial heterogeneity, the faster it fades. Further, the speed of mixing is dependent upon the supernovae (SN) rate.

In the present work, we numerically simulate the galactic chemical evolution of the SLRs, $^{26}$Al, $^{36}$Cl, $^{41}$Ca, $^{53}$Mn, and $^{60}$Fe over the entire radial extent of the Milky Way galaxy from 2-18 kpc from the galactic center to understand the temporal and spatial distribution of the SLRs. We could specifically decipher the stellar contribution of SLRs from low and intermediate-mass stars and supernova type II and Ib/c during the galactic chemical evolution. Further, using the production rate of SLRs we make an attempt to understand the context of the birth of the solar system within a stellar cluster as a natural outcome of the GCE model. This is studied by considering a heterogeneous GCE model that simulates the galaxy over distinct spatial grids as a sequence of formation and evolution of several generations of stellar clusters. The formation of the individual stellar clusters/associations is studied in the solar neighborhood to understand the origin of the solar system and the initial abundance of SLRs, especially, $^{26}$Al, $^{41}$Ca, $^{60}$Fe and $^{53}$Mn. The numerical techniques involved in the simulations are presented in section 2. The results of the various models are mentioned in section 3, and discussed in section 4. Finally, the major conclusions drawn from the present work are mentioned in section 5.

## 2. Numerical Simulations

The basic approach adopted in the present work is based on the evolution of the Milky Way galaxy in terms of the formation and evolution of stars in stellar clusters. The major constraints for a GCE model are imposed by the predicted star formation rate (SFR), the supernovae rates, the surface gas mass density, the surface stellar mass density and the observed elemental abundance distribution of the halo, thick and thin disc stars of the galaxy. The well-observed solar neighborhood imposes stringent constraints on the evolution of the galaxy in the GCE models. In order to deduce the stellar nucleosynthetic contribution of SLRs, a combination of homogeneous and heterogeneous GCE numerical models for the evolving galaxy and the solar neighborhood are presented here. The preliminary results of this work have been presented in a recent international conference (Kaur & Sahijpal 2019).

Based on our recent numerical simulation work on the evolution of the galaxy (Sahijpal & Kaur 2018), the galaxy is radially divided into eight annular rings of 2 kpc width each at a distance of 2-18 kpc from the center of the galaxy. In this approach based on N body Monte Carlo technique, the galaxy accretes intergalactic gas according to an assumed exponential infall rate. Inside each annular ring, the numerous generations of stars in the mass range 0.1-100 $M_\odot$ form and evolve in the galaxy according to the prescribed star formation rate (SFR) and initial mass function (IMF). The formed stars evolve over their mass and metallicity dependent life spans, and subsequently enrich the ISM by ejecting freshly synthesized elements. In this manner, the metallicity of ISM gradually increases and the next generation of stars are formed from the enriched ISM gas. In the simulations, the solar system is assumed to have formed 4.56 Gyr ago within the 4$^{th}$ annular ring which is commonly referred to as the solar neighborhood. The solar annular ring is assumed to have acquired a metallicity of ~0.0143 (Asplund et al 2009) and [Fe/H] ~0 at the time of the formation of the solar system. The exact value of the solar metallicity even now remains a debatable issue. The results presented in the present work will not be substantially influenced by the exactness in the solar metallicity value that has been considered to be in the range of 0.014-0.019. Further, we have not considered the anticipated radial migration of sun subsequent to its formation as this will not substantially influence the major inferences drawn from the present work.

The GCE simulations were performed to predict the abundance of SLRs in the solar system along with the stable nuclides. As discussed in the previous section the SLRs could have been synthesized by massive stars formed prior to the formation of the solar system. In order to explain the contribution of SLRs in the solar system, we have performed simulations for two sets of GCE models. This includes the homogeneous (*Homo-*) and the heterogeneous (*Heter-*) GCE models. In the homogeneous GCE model, the entire galaxy is evolved in terms of individual evolution of the annular rings of radial width 2 kpc each across the 2-18 kpc radial extent of the galaxy. The nucleosynthetic ejecta of stars subsequent to their evolution are instantaneously homogenized across the entire annular ring. These models provide a gradual homogeneous evolution of the galaxy for SLRs in a manner identical to the stable nuclide evolution.

We performed the simulations for the homogeneous evolution of SLRs by choosing the Model H from our earlier work on the stable isotopic evolution of the galaxy (Sahijpal & Kaur 2018). We adopted an identical approach as far as the evolution of the galaxy and the stellar nucleosynthetic prescription is concerned. We avoid a detail discussion of this numerical simulation and present a brief summary of its various features. The Model H deals with three infall accretion scenario for the galaxy formation (Micali, Matteucci & Romano 2013). It explains the observed elemental abundance evolutionary trends of the halo, the thick and thin disc stars in a better manner compared to two infall accretion models. The galactic halo and thick disc are formed within the initial 0.2 Gyr and 1.25 Gyr, respectively. The thin disc forms according to the inside-out criteria, with the solar neighborhood accreting mass over a time-scale of ~7 Gyr. The inner regions of the thin disc formed earlier and the outer regions formed later according to time-scale which is a function of distance from the galactic center. This inside-out criterion contributed to the formation of abundance gradients in the galactic disc (Chiappini et al. 1997, 2001; Micali et al. 2013; Sahijpal & Kaur 2018). The value of halo surface mass density is considered to be 17 $M_\odot$ pc$^{-2}$ within the inner 10 kpc, and thereafter, it is assumed to vary with r$^{-1}$ dependence, where 'r' is the radial distance from the center. The thick disc surface mass density is assumed to be 24 $M_\odot$ pc$^{-2}$ up to 10 kpc, and thereafter, its contribution is considered to be 40 percent of the total disc surface mass density. The total surface mass density is assumed to be 64 $M_\odot$ pc$^{-2}$ in the solar ring located at 8-10 kpc. Even though we could perform the simulations for GCE of SLRs for our other models dealing with inter-annular ring gas mixing (Sahijpal & Kaur 2018), we have avoided these models due to the limited anticipated gas mixing over the time-scales comparable to the half-lives of the radionuclides. The simulation stars are formed according to the prevailing star formation rate and the initial mass function (Scalo 1998; Matteucci 2003, Sahijpal & Kaur 2018). The stars are formed with discrete masses of 0.1, 0.4, 0.8, 1, 1.25, 1.75, 2.5 $M_\odot$ and integer masses from 3-100 $M_\odot$ (except 9, 10 $M_\odot$) following a power-law distribution. We have not considered the stars with masses 9 and 10 $M_\odot$ because the evolution of these stars is uncertain due to the uncertain mass-losses during their evolution. Further, the nucleosynthetic yields of these stars are also not available (see e.g., Sahijpal & Gupta 2013). The slope of the power-law is defined in such a way that the low mass stars are more abundant than the massive stars. The value of the exponent is considered to be 0 for the mass range, 0.1-1 $M_\odot$ and 1.7 for 1-100 $M_\odot$ stars. The IMF is obtained in such a manner that the mass fraction contribution of stars in the mass range of 0.1-0.8 $M_\odot$ and 1-100 $M_\odot$ towards the integrated IMF comes out to be ~0.45 and ~0.55, respectively.

All the simulations were performed with a time step of 1 million years (Myr.) (Sahijpal & Kaur 2018). The nucleosynthetic yields of the stable and SLRs for the majority of the stars in the mass range 0.1-100 $M_\odot$ are used to access the production rates of SLRs in the entire galaxy. This includes the supernovae SN II (Wooley & Weaver 1995, Limongi & Chieffi 2018), supernovae SN Ia (Iwamoto et al. 1999) and AGB stars (Karakas & Lattanzio 2003; Karakas 2010) contributions. Except for $^{26}$Al stellar contributions for the low and intermediate stars (Karakas 2010), we could not manage to obtain the stellar contributions of other SLRs for AGB stars. We rather used the FRUITY database[1] (see e.g., Cristallo et al. 2015) to determine the SLR contributions ($^{41}$Ca, $^{36}$Cl & $^{60}$Fe) of these AGB stars across all the metallicity models. In order to understand the influence of the nucleosynthesis of massive stars on GCE, we have compared different models with nucleosynthetic yields of massive stars from Woosley & Weaver (1995) (WW95) and Limongi & Chieffi (2018) (LC18). A preliminary comparison of the GCE models developed on the basis of the WW95 stellar nucleosynthetic yields and Limongi & Chieffi (2012) has been earlier made by Sahijpal (2014b). The details of WW95 yields are discussed by Sahijpal & Gupta (2013). The LC18 provide yields for stars having initial masses, 13, 15, 20, 25, 40, 60, 80, and 120 M$_\odot$ for rotational velocity of 0, 150 and 300 km s$^{-1}$ and initial metallicity [Fe/H] = 0, -1, -2, and -3. We have considered non-rotating and rotating cases with the rotational velocities of 0 and 300 km s$^{-1}$, respectively. The yields are interpolated and extrapolated to obtain the yields for stars in the mass range 11-100 M$_\odot$. We performed four sets of *Homo*-GCE model simulations with a distinct set of stellar nucleosynthetic yields. The details are mentioned in table 1. In Model I, the core-collapse supernova data was taken from WW95. The AGB star data (including $^{26}$Al) was taken from Karakas 2010 (K10) and the yields of $^{41}$Ca, $^{36}$Cl & $^{60}$Fe of AGB stars were taken from the FRUITY database. It should be mentioned here that in general there is a limited justification in using the stellar yields obtained by two different groups for different isotopes even for the stellar models with the same metallicities and masses. However, in the present circumstance, it is not possible for us to obtain the yields of all the SLRs for the AGB stellar models by Karakas (2010). In Model II, the core-collapse supernova data was taken from WW95. The AGB star data was taken from K10 and the yields of $^{26}$Al, $^{41}$Ca, $^{36}$Cl & $^{60}$Fe of AGB stars were taken from the FRUITY database. The $^{26}$Al yields of AGB stars, especially, the 4-6 M$_\odot$ intermediate-mass stars are higher in the case of K10 models compared to the FRUITY database by up to two orders of magnitude for the 0.008 metallicity stellar models that substantially determines the galactic chemical evolution around the formation time of the solar system with an assumed solar metallicity value of ~0.0143. The stellar yields with the metallicity in the range of 0.008 - 0.02 are estimated on the basis of interpolation of the yields from the models with the available metallicities, 0.008 and 0.02. Based on the observed differences in the $^{26}$Al yields from the two AGB databases, we anticipate identical differences among the yields obtained from the two models for the other SLRs. However, the associated stable nuclides yields do not differ significantly among the two models. We made a comparison of the stable isotopic yields among the two nucleosynthetic models for the stable nuclides, $^{27}$Al, $^{40}$Ca, $^{35}$Cl & $^{56}$Fe in the case of stellar models with metallicity values of 0.008 and 0.02. The yields differ up to a factor of 2 in the case of $^{35}$Cl, and by a factor of 1.5 in the case of $^{27}$Al, $^{40}$Ca, & $^{56}$Fe. However, we do anticipate uncertainties in the SLR yields based on the two model that can influence the GCE of SLRs from the AGB stars contribution. In Model III, the core-collapse supernova yield data was taken for the zero rotational velocity stars from LC18. The AGB star data (including $^{26}$Al) was taken from K10 and the yields of $^{41}$Ca, $^{36}$Cl & $^{60}$Fe of AGB stars were taken from the FRUITY database. In the Model IV, the core-collapse supernova data was taken for 300 km s$^{-1}$ rotational velocity stars from LC18. The AGB star data (including $^{26}$Al) was taken from K10 and the yields of $^{41}$Ca, $^{36}$Cl & $^{60}$Fe of AGB stars were taken from the FRUITY database. It should be mentioned that we could not succeed in acquiring any additional data except for the stellar nucleosynthetic data considered above.

---

1  http://fruity.oa-teramo.inaf.it/

The fraction of the stellar population in the mass range of 3–8 M$_\odot$ and 11–16 M$_\odot$ can evolve as binary stars and eventually evolve as SN Ia. The details of the synthesis of the progenitor stellar population for SN Ia were adopted from Sahijpal & Kaur (2018). A model with a delay time distribution (DTD) based upon single degenerate (SD) and double-degenerate (DD) are explored to access the SN Ia rates (Matteucci et al. 2009).

Although the homogeneous GCE model, in the present and earlier works, provides an averaged gradual production rates of the SLRs during the evolution of the galaxy, the anticipated assumption of the SLRs homogenization over the entire annular ring subsequent to stellar ejecta is not possible due to the extremely short half-lives of the nuclides compared to the mixing time-scales. In order to relax this drastic large-scale homogenization assumption, we have performed simulations for heterogeneous GCE models (*Heter-*) where the homogenization of the nucleosynthetic ejecta from stars occur over a localized region rather than the entire annular ring.

The heterogeneous GCE models (*Heter-*) were numerically simulated for the solar annular ring alone. The features developed by Sahijpal (2013) for the heterogeneous GCE models were integrated with the GCE formulation recently developed by Sahijpal & Kaur (2018) to understand the production of the SLRs. We ran several simulations with the *Heter*-GCE models corresponding to WW95 and LC18 stellar nucleosynthetic yields (table 1). The Models AI-EI are based on the data taken from WW95 and K10, whereas, the Models DIII and DIV are based on the data taken from LC18 and K10 for the stellar rotational velocities of 0 and 300 km s$^{-1}$, respectively. In these models the solar annular ring (8-10 kpc) is further divided into independent 100-800 spatial grids of area ~1.13-0.14 kpc$^2$, depending upon the model. These spatial grids evolve distinctly in terms of the chemical evolution of the galaxy (Sahijpal 2013). Rather than evolving all the 100-800 grids, we evolve only 8 distinct grids that provide a suitable representation of a large ensemble of grids in terms of diversity. The star formation occurs discretely in episodes of independent stellar clusters within each isolated grid according to the prevailing star formation rate (SFR) and IMF with a temporal gap of over several tens to hundred million years depending upon the model (table 1). This temporal gap provides a time interval before which the next generation of star formation is ready to commence from the newly formed and evolved molecular cloud within the grid. This temporal gap is determined based on a random number generation. The majority of the simulations were run with the formation of a typical stellar cluster over a time duration of 1 Myr that also serves as a time step for the simulations. However, we ran two simulations with the Models CI & DI where a single stellar cluster is assumed to form over a time-span of 3 Myr and 10 Myr, respectively. In the second case, the star formation within a single stellar cluster occurred with a gap of 1 Myr after each 2 Myr of star formation. It should be mentioned that we are presenting here simulations with a limited possible parametric combination. It is possible to work out simulations with an alternative set of parameters that include extending the star formation time-scale within a single stellar cluster.

The mass of a stellar cluster formed at any epoch is evaluated on the basis of a random spread in the average mass inferred according to the prevailing star formation rate (SFR) (Sahijpal 2013). The average mass of a simulation stellar cluster along with the range in the mass at the time of formation of the solar system is presented in table 1 for the various models. The synthesized massive stars (>11 $M_\odot$) due to their short life eject their nucleosynthetic yields within the associated grid in which these stars are formed, thereby, restricting the spatial region for homogenization of the stable nuclides and SLRs. The localized mixing of ejecta with ISM (interstellar medium) gives rise to the heterogeneous nature of the evolution. The nucleosynthetic contributions of the SLRs are dominated by the massive stars because of the localized mixing in the spatial grid. The nucleosynthetic yields from the low-mass AGB stars and SN Ia are not considered in the *Heter*-GCE models as far as the SLRs are concerned.

There are two critical issues that are relevant regarding the *Heter*-GCE models for the proper interpretation regarding the production and homogenization of SLRs. These issues are, i) the average (and the range in) mass of the stellar cluster/association, and ii) the extent of the spatial homogenization of SLRs with ISM during their lifetimes. The deduced star formation rates (SFR) in the case of models AI-EI & DIII-DIV (table 1) indicate an average mass of stellar clusters to be in the range of ~$3\times10^4$ $M_\odot$ to $3\times10^5$ $M_\odot$ around the time of the formation of the solar system. The observed total mass range, however, varies over three orders of magnitude, i.e., in the range of $10^{3-5}$ $M_\odot$. Hence, the estimated average mass could be up to two orders of magnitude high compared to the observed estimates even though stellar associations with these masses do exist (Lada & Lada 2003; Lada 2010; Lada et al. 2010; Murray 2011). The observed stellar clusters exhibit a power-law distribution in mass with an upper limit of ~$10^5$ $M_\odot$. We introduce a reducing parameter, '$\eta$' (in the range of ~1-100), in order to anticipate the behavior of SLR production with a reduced average mass of a typical stellar association/cluster. This would result in an identical reduction in star formation rates. This, in turn, will linearly reduce the supernovae rates, the nucleosynthetic yields of SLRs and their deduced ratios compared to those anticipated for the models (table 1). However, this reduction in the SLR abundances can be appropriately compensated by reducing the spatial extent of the homogenization of SLRs within the grid having an area of ~0.14-1.13 kpc$^2$. We introduce a parameter, 'χ' (in the range of ~$10^0$-$10^{-2}$) that would substantially reduce the tangible homogenization area of the massive stars produced SLRs within a grid. A reduction in the homogenization area by a factor of up to 100 will confine the SLR contamination region to be right within the vicinity of massive stars. It should be noted that the parameters, η, and χ, are crucial for SLRs. The two parameters are related to the mass of the stellar cluster and determines the spatial extent of SLR homogenization in a localized region. These parameters do not substantially influence the stable nuclides GCE, except for the production of spatial heterogeneities that are homogenized over longer time-scales >$10^8$ years (Sahijpal 2013), thereby, maintaining a steady-state evolution of metallicity within the galactic annular ring.

### 3. Results

Our simulation results predict the average abundance distribution evolution of SLR over the entire Milky Way galaxy. Further, we also provide high spatial resolution SLR production trends in the solar neighborhood during the entire evolution of the galaxy with the major emphasis at the time of the formation of the solar system in order to understand the associated astrophysical environment. The *Homo*-GCE model predicts the average abundance evolution of SLRs, $^{26}$Al, $^{36}$Cl, $^{41}$Ca, $^{53}$Mn and $^{60}$Fe, on account of homogenous GCE for the eight annular rings from 2-18 kpc over the galactic time scale. We have performed simulations with four *Homo*-GCE models with nucleosynthetic yields from different sources (table 1). The predicted trends are presented in figs. 1, 2, 9 & 10 (figs. 9 & 10 suppl. data). The predicted abundance distribution of SLRs around the time of the formation of the solar system as determined by the *Heter*-GCE models are shown in figs. 3-7 & 11-12 (figs. 11 & 12 suppl. data). The results for these models are presented in the form of a sequel of consecutive formation of stellar clusters within a single spatial grid. The formation of the solar system could have occurred in association with one of these stellar clusters presented in figs. 3-7 & 11-12, depending upon the contribution of the SLRs.

In the case of *Homo*-GCE models, the figs. 1a, 2a, 9a & 10a, represents the abundance evolutionary trends of $^{26}$Al/$^{27}$Al, $^{36}$Cl/$^{35}$Cl, $^{41}$Ca/$^{40}$Ca, $^{53}$Mn/$^{55}$Mn and $^{60}$Fe/$^{56}$Fe, within the solar annular ring, 8-10 kpc, over the galactic time-scale. The experimentally determined initial abundance ratios of SLRs in the ESS with respect to their stable nuclides (table 2) are also marked in the figures for comparison with the GCE predictions. The evolution of the absolute abundances of SLRs, $^{26}$Al, $^{36}$Cl, $^{41}$Ca, $^{53}$Mn, and $^{60}$Fe are presented in figs. 1b, 2b, 9b & 10b, for the 8-10 kpc solar annular ring. The abundance of all the SLRs increases very sharply during the initial 0.5 Gyr at the time of the formation of the galactic halo and subsequently stabilizes within ~2 Gyr. During the initial 2 Gyr, the star formation rate (SFR) is high due to the rapid accretion of extragalactic matter. This is followed by high supernova (SN Ib/c and II) rates. The massive stars with short life-time evolve fast and contribute SLRs rapidly to the interstellar medium. The SFR reduces at the end of the halo and thick disc formation episodes. With the decline in the SFR, the SN (II, Ib/c) rates also reduce during the formation of halo and the thick disc which in turn reduces the SLR production. Table 2 presents the GCE deduced SLRs ratios at the time of the formation of the solar system for the various models in comparison with the most widely accepted reference values of SLRs in the ESS.

The production rates of SLRs follow the star formation rate (SFR) trends (Sahijpal & Kaur 2018). Due to the dip in SFR at the end of the halo-thick disc phase, the abundance of all the SLRs declines sharply afterward and subsequently evolves to achieve the present-day values. The abundance ratios of the two most well-studied SLRs with respect to their stable isotope, $^{26}Al/^{27}Al$ and $^{60}Fe/^{56}Fe$, are presented in figs. 1c, 2c, 9c & 10c, across the entire galaxy for all galactic annular rings. The initial reference abundance ratios of the two SLRs in the ESS with respect to their stable nuclides are also marked in the figures by horizontal bars for comparison with the GCE predictions. The evolution of the absolute abundances of $^{26}Al$, $^{36}Cl$, $^{41}Ca$, $^{53}Mn$ and $^{60}Fe$ over the galactic time scale for the entire galaxy are presented in figs. 1d-h, 2d-h, 9d-h & 10d-h.

The results of the seven sets of *Heter*-GCE Models, AI, BI, CI, DI, EI, DIII, and DIV are presented in figs. 3-7 & 11-12. The details of the model parameters are provided in table 1. Various feasible astrophysical scenarios for the formation of the solar system in association with a stellar cluster(s) were simulated and the best ones are studied in details. In our simulations, the solar system formed as a consequence of the Milky Way galaxy evolution over the galactic time scale approximately ~4.56 Gyr ago. Further, we have explored the various configurations of the stellar clusters/association formed around the time of the solar system formation. The observational data from the study of various meteoritic samples indicates the need for a local stellar source(s) for the existence of SLRs in the solar nebula.

The star formation rates and supernovae rates for the *Heter*-GCE models are presented in figs. 3a-7a & 11a-12a and figs. 3b-7b & 11b-12b, respectively, for one of eight randomly selected spatial grids whose evolution is observed over the galactic time scale. The deduced ratios of SLRs with respect to their most stable isotopes, $^{26}Al/^{27}Al$, $^{36}Cl/^{35}Cl$, $^{41}Ca/^{40}Ca$, $^{53}Mn/^{55}Mn$ and $^{60}Fe/^{56}Fe$ are presented in figs. 3c-7c & 11c-12c along with the temporal evolution of the absolute abundance of $^{26}Al$, $^{36}Cl$, $^{41}Ca$, $^{53}Mn$, and $^{60}Fe$, in figs. 3d-7d & 11d-12d. The contribution of massive stars (>11 M☉) from the stellar cluster is only considered for the SLRs yields as the SN Ia and AGB stars steady-state galactic contributions are substantially low compared to the localized spiked contributions from the massive stars. Due to the low mass of their progenitor stars, the AGB stars and SN Ia evolve over longer time durations compared to massive stars, and hence, substantially migrate across the galaxy, thereby, providing a gradual steady-state enrichment of stable nuclides to the galaxy.

In general, we limit the presentation of results to a single spatial grid for each model in order to avoid repetition. However, it should be noted that the result from the other spatial grids infers almost identical results. This is demonstrated in the metallicity (Z) and [Fe/H] evolutionary trends for the Model BI, presented in figs. 8a & b, respectively, for eight distinct spatial grids. The Z and [Fe/H] evolutionary trends for these eight grids, randomly selected from the solar annular ring, yield almost identical results despite the spatial heterogeneities.

## 4. Discussion

One of the major objectives of the present work is to understand the origin and evolution of the distribution of SLRs in the entire Milky Way galaxy on account of the galactic chemical evolution over the galactic time-scale. The average abundance distribution evolution of SLRs over the entire galaxy provides a detailed account of the star formation history of the galaxy. The average *Homo*-GCE trends of SLRs follow the star formation rate. In general, the average GCE trends show that the production of SLRs is high in the inner regions of the galaxy and gradually reduces towards the outer regions on account of comparatively low star formation (fig. 1). The reduction in the abundance is also observed at all the distances over the evolution of the galaxy. It should be noted that the recent work (Fujimoto et al. 2018) has also proved that there should be a strong co-relation in the abundance distribution trends of $^{26}$Al and $^{60}$Fe with the galactic star formation rate.

In the case of four *Homo*-GCE models, as shown in figs. 1a, 2a, 9a, 10a, the inferred interstellar medium values of $^{26}$Al/$^{27}$Al, $^{36}$Cl/$^{35}$Cl, $^{41}$Ca/$^{40}$Ca, $^{53}$Mn/$^{55}$Mn and $^{60}$Fe/$^{56}$Fe in the solar annular ring at the time of the formation of the solar system are listed in table 2. These estimates have to be compared with the measured initial reference ratios in the ESS phases. The *Homo*-GCE model predicted ratios shown in table 2, falls short by an order of magnitude in explaining the initial $^{26}$Al/$^{27}$Al and $^{36}$Cl/$^{35}$Cl values in the ESS, whereas, the predicted GCE values of $^{41}$Ca/$^{40}$Ca and $^{53}$Mn/$^{55}$Mn are quite high compared to the observed values in the meteoritic phase. The GCE prediction of $^{60}$Fe/$^{56}$Fe is high compared to the ESS value by two orders of magnitude across all the models irrespective of the stellar nucleosynthetic prescription. This could explain possible GCE contribution of $^{60}$Fe to the ESS.

There is no proposed mechanism of enhancing the GCE $^{26}$Al/$^{27}$Al and $^{36}$Cl/$^{35}$Cl in order to explain the origin of $^{26}$Al and $^{36}$Cl in the ESS. The reduction in the GCE contribution of $^{41}$Ca/$^{40}$Ca to the ESS can be achieved by appropriately increasing the free decay time interval between the last stellar nucleosynthetic event that provided the $^{41}$Ca stellar inventory and the formation of the earliest solar system phases that entrapped the radionuclide. However, this scenario will not work in the case of $^{53}$Mn due to its long lifetime. The only plausible mechanism for the reduction in $^{53}$Mn is by reducing its SN II yields that do not seems to work based on the two distinct sets of stellar yields of WW95 and LC18.

There are systematic differences among the GCE predictions from models based on distinct set of stellar yields (table 2). The GCE predictions of Model IV based on LC18 for rotating stars with velocities of 300 km s$^{-1}$ infer higher values compared to WW95 models and the LC18 model with non-rotational massive stars. As mentioned earlier, the $^{26}$Al yields of K10 are higher than the yields obtained from the FRUITY database for the intermediate-mass stars. This results in higher GCE production of $^{26}$Al in Model I compared to the Model II.

The two short-lived radionuclides, $^{26}$Al and $^{60}$Fe, have been also monitored by gamma-ray telescopes for their abundance in the interstellar medium. The gamma-ray measurements from the INTEGRAL satellite were used to estimate the flux value for $^{60}$Fe. The inferred ratio of the measured fluxes yielded a value of 0.148 for $^{60}$Fe/$^{26}$Al. This flux ratio infers a value of 0.31 for $^{60}$Fe/$^{26}$Al (Wang et al. 2007; Huss et al. 2009; Fujimoto et al. 2018). The flux is translated to the abundance by considering the amount of stable isotope and the mass of the gas in the galaxy. After scaling the present-day observed gamma-ray fluxes to the time of formation of the solar system, the inferred values for $^{26}$Al/$^{27}$Al and $^{60}$Fe/$^{56}$Fe are 4.0×10$^{-6}$ and 4.4×10$^{-8}$, respectively (Diehl et al. 2006a, b; Wang et al. 2007; Huss et al. 2009; Sahijpal 2014a). Our results almost match as far as $^{26}$Al is concerned, especially, for the Models I, III & IV (table 2; figs. 1a, 2a, 9a, 10a). However, our predictions of $^{60}$Fe are high by a factor of ~2-3 compared to the observed value for the various models (table 2). The present interstellar isotopic ratios from gamma-ray flux for $^{26}$Al/$^{27}$Al and $^{60}$Fe/$^{56}$Fe are 8.4×10$^{-6}$ and 2.7×10$^{-7}$, respectively (Huss et al. 2009). These values have to be compared with our present-day predictions given in table 2. $^{26}$Al/$^{27}$Al is low up to a factor of 4, and $^{60}$Fe/$^{56}$Fe is low up to a factor of 3.

The GCE abundance ratios of $^{26}$Al/$^{27}$Al and $^{60}$Fe/$^{56}$Fe are higher in the inner regions of the galaxy compared to the outer regions as shown in figs. 1c, 2c, 9c & 10c because of the higher SFR in the inner regions due to fast accretion of gas. The absolute abundance of $^{26}$Al in our galaxy has also been inferred. The gamma-ray flux calculations provide an estimate of the total amount of $^{26}$Al in the entire Milky Way galaxy to be 2.0 ± 0.3 M$_\odot$ (Diehl et al. 2010, Diehl 2016) and 1.7 ± 0.2 M$_\odot$ (Martin et al. 2009). Our absolute predictions for $^{26}$Al are low by an order of magnitude for the assumed planar geometry of our GCE model based on annular rings. As far as $^{26}$Al is concerned, we have found that the AGB stars and massive stars contribute almost equally to the steady-state GCE of the radionuclide as far as Model I is concerned that is based on the $^{26}$Al yields of K10 rather than the FRUITY database. The AGB stars and SN Ia provide a steady-state contribution, whereas, the contributions from massive stars come in the form of localized enhancements.

The other major objective of the present work is to understand the possibility of the formation of the solar system in association with stellar cluster(s) in context with the heterogeneous evolution of SLRs in the galaxy. This is achieved by developing a heterogeneous GCE model for SLRs by taking into account the fact that SLRs from stellar sources, especially, the massive stars, cannot homogenize over large spatial extents of the galaxy within their short life spans. The *Heter*-GCE of independent spatial grids with an area of 0.14 -1.13 kpc$^2$ is studied. These models indicate that the formation of the solar system from a molecular cloud along with another stellar cluster (s) can explain the observed distribution of the SLRs in the ESS. This scenario seems to be compatible with the gradual evolution of the galaxy in terms of sequel formation and evolution of stellar clusters. The contributions of SLRs to the ESS need not be confined to a single stellar source. The observed abundance distribution could be a collective result of various sources that include massive Wolf-Rayet star(s), SN II and SN Ib/c.

As mentioned earlier, in all *Heter*-GCE models (figs. 3-7, 11, 12), the main contribution of the SLRs comes from massive stars in terms of localized enhancements in the SLRs yields. This fact is also supported by the observed irregular distribution of $^{26}$Al along the plane of the galaxy as deciphered from the gamma-ray observations that indicate the role of massive stars, formed inside the clusters, as the major contributors (Prantzos & Diehl 1996).

The figs. 3-7a, 11a, 12a present the SFR in one of the randomly selected spatial grids for the *Heter*-GCE models. The SFR for a time interval of ~9.18-9.50 Gyr around the formation of the solar system is shown in further detail in the inset figure. The temporal gap between the consecutive episodes of the formation of the stellar clusters is chosen randomly from the temporal range mentioned in table 1 for the various models. The figs. 3-7b, 11b, 12b present the deduced supernovae (SN II, SN Ib/c and SN Ia) rates. The SN II and SN Ib/c rates correlate with the star formation rates in the figs. 3-7a, 11a, 12a. The ratios of the SLRs with respect to their stable isotope and their absolute abundances for the distinct *Heter*-GCE models are presented in figs. 3-7c, 11c, 12c and figs. 3-7d, 11d, 12d, respectively. On contrary to the *Homo*-GCE model, the *Heter*-GCE models are capable of producing high SLRs yields over localized regions (≤ 1 kpc$^2$) during the initial stages of the evolution of stellar associations /clusters. The maxima achieved for SLR in some cases is high enough to explain the canonical value of $^{26}$Al/$^{27}$Al in the ESS. However, these contributions are directly associated with the evolution of massive stars within a localized stellar cluster(s) at a particular instant. The scenario for the contribution of $^{26}$Al by stellar cluster(s) to a nearby presolar cloud at the required canonical level of $^{26}$Al/$^{27}$Al in ESS seems to be an intriguing scenario provided the stellar contributions of other SLRs, $^{60}$Fe and $^{53}$Mn can also be explained in a self-consistent manner. However, the inferred yields of $^{53}$Mn/$^{55}$Mn and $^{60}$Fe/$^{56}$Fe are more than two orders of magnitude high compared to their ESS initial reference values (table 2). In the following, we propose a scenario that could explain in a self-consistent manner the abundance of the majority of the SLRs considered in this work. As mentioned earlier, since the stars are mostly formed in a stellar cluster, the formation of the solar system has to be reconciled with the association of stellar cluster(s), and probable the SLR contributions from these clusters. Before exploring into a self-consistent scenario for the origin of SLRs within the present theoretical framework we explore the nature of the range in the mass of stellar clusters and the spatial scale for the homogenization of stellar ejecta from the evolved stars.

In the present work, the average mass of the stellar clusters/associations formed in the grids around the time of the formation of the solar system is $10^4$-$10^5$ $M_\odot$. The total mass range can vary up to three orders of magnitude (Table 1). The astronomical observations of the star-forming regions suggest an average mass of the cluster/association to be $10^3$-$10^4$ $M_\odot$ (Lada & Lada 2003; Lada et al. 2010; Murray 2011). There are, however, observational evidence for the existence of even more massive stellar clusters as in the present case. As mentioned in section 2 in order to understand the implications of the choice of a low mass stellar cluster on the production of SLRs, we can reduce the average stellar cluster mass by scaling down the mass by a parameter 'η', in the range of 1-100. This approach circumvents the short-coming of our models to simulate with smaller spatial grid sizes. The main reason for the choice of the adopted massive stellar clusters/association in our model is the limitation on the grid size. The grid size in our work is large enough to host multiple star-forming regions in a contemporary manner. The average radius of a GMC can be ~30 pc (Tasker & Tan 2009), and the average grid size in our simulations is 0.42 kpc$^2$ that can harbor multiple molecular clouds. The results for the simulation with the smallest grid of 0.14 kpc$^2$ are presented in fig. 12. If we consider a two-dimensional geometry, the equivalent radius of the GMC will be 0.32 kpc for the smallest grid size of 0.14 kpc$^2$ opted in our simulations. As mentioned earlier (Tasker & Tan 2009; Murray 2011; Fall & Chandar 2012), the average mass of gas contained inside the GMC is $10^6$-$10^7$ $M_\odot$ and the conversion efficiency of gas into stars is 0.08. This will result in a stellar mass of $10^5$-$10^6$ $M_\odot$ inside a GMC. A single GMC contains multiple star-forming regions in the form of distinct stellar clusters.

The reduction in the average stellar cluster mass by a factor η will proportionally reduce the stellar production of SLRs from the massive stars from the values anticipated in the figs. 3-7c, 11c, 12c and figs. 3-7d, 11d, 12d. This reduction in the SLRs yields can, however, be compensated by introducing another simulation parameter 'χ', as a multiplier of η, such that it reduces the geometrical dilution factor responsible for the homogenization of SLRs over distinct spatial extents. The value of 'χ' can be anticipated in the range of 0.01-1. Thus, the factors, η, and χ, determine the mass of the stellar cluster and the extent of spatial homogenization of SLRs subsequent to their ejection from massive stars.

On the basis of the *Heter*-GCE models AI-EI, DIII & DIV, and the observations of the star-forming regions, we propose a hypothesis for the self-consistent origin of the SLRs. Based on the deduced SLR trends in figs. 3-7c, 11c, 12c and figs. 3-7d, 11d, 12d, we hypothesize that the formation of the solar system occurred subsequent to the final stages of the evolution of a stellar cluster (say, Cluster-A). The onset of the formation of the stellar cluster-A could have initiated ≥ 25 Myr. prior to the formation of the solar system in another cluster-B, probably formed within the same giant molecular cloud complex. The stars in the cluster-A evolved through core-collapse supernovae stages and contaminated the local ISM with their stellar nucleosynthetic debris that also contained $^{53}$Mn and $^{60}$Fe. The presolar molecular cloud probably acquired the requisite amount of the initial $^{53}$Mn/$^{55}$Mn and $^{60}$Fe/$^{56}$Fe values in the ESS (table 2) from the cluster-A stars. The three thick grey dashed lines in fig. 3c indicates three possible alternative manners by which the presolar molecular cloud could have acquired the two SLRs inappropriate abundance. The parameters, η, and χ, dealing with the stellar cluster-A mass and the extent of the spatial homogeneity associated with the stellar ejecta of cluster-A stars will determine the time-scale between the formation of the solar system after the commencement of the formation of the stellar cluster-A. This scenario provides a natural mechanism to explain the ESS initial reference values of $^{60}$Fe and $^{53}$Mn. In fact, the homogeneous GCE contribution of SLRs, especially, $^{60}$Fe due to their short life, in reality, would eventually come from the finally formed and evolved stellar cluster(s). As mentioned in section 1, the spatial heterogeneities existing over a small scale of <100 pc can be homogenized within a couple of million years by turbulent currents and stellar winds. Hence, the cluster-A could be several tens of parsecs away from the ISM that eventually evolved into the parent molecular cloud for the cluster-B, the host cluster of the sun.

Once the $^{60}$Fe and $^{53}$Mn contributions to the presolar molecular cloud is appropriately explained from the cluster-A stars, the $^{60}$Fe contributions from the stellar cluster-B have to be strictly curtailed from any core-collapse supernovae contribution. The stellar cluster-B, host to the sun, was probably having low mass. The low stellar cluster mass could have hampered the formation of massive stars, thereby, restricting the localized high $^{60}$Fe contributions of core-collapse supernovae from the same cluster. Alternatively, the presolar molecular cloud gravitationally collapsed prior to the significant injection of supernova $^{60}$Fe rich ejecta.

The stellar contribution of $^{26}$Al (and probably $^{41}$Ca) to ESS in the proposed scenario would require the role of an independent Wolf-Rayet wind (Sahijpal & Soni 2006; Dwarkadas et al. 2017) without any trailed $^{60}$Fe rich supernova contribution. Sahijpal & Soni (2006) considered the role of a non-rotating and a rotating single 60 M$_\odot$ Wolf-Rayet star in the injection of $^{26}$Al into the presolar molecular cloud. The associated dynamics and the time-scales related to the injection, and the anticipated stable isotopic effects in CAIs in ESS were deduced. It was demonstrated that the Wolf-Rayet ejecta from a massive star at a distance of 5-25 pc from the presolar molecular cloud can explain the abundance of $^{26}$Al (and probably $^{41}$Ca) in the ESS without invoking large stable isotopic anomalies in the ESS phases. The production of $^{36}$Cl falls short by a factor of 6 in this scenario compared to the ESS abundance. Since the anticipated time for the contribution of $^{26}$Al from Wolf-Rayet star to ESS is less than one million years, the massive star formed and evolved in a distinct stellar cluster (say cluster-C) at a distance of up to ~25 pc from the presolar molecular cloud. As the massive Wolf-Rayet stars are the earliest to evolve within a stellar cluster, the formation of the cluster-C could have commenced ~3-4 Myr. prior to the formation of the solar system if we invoke a 60 M$_\odot$ Wolf-Rayet star. The recent work by Dwarkadas et al. (2017) considered a detailed dynamical evolution of a Wolf-Rayet bubble and the possible contamination of presolar molecular cloud in terms of SLR contribution. In the present scenario, the three thick grey dashed lines in fig. 3c indicates three possible alternative manners by which the presolar molecular cloud acquired $^{60}$Fe and $^{53}$Mn from the local ISM and freshly synthesized $^{26}$Al (and $^{41}$Ca) from Wolf-Rayet wind. The latter is represented in the figure as a spiked contribution.

The proposed hypothesis involving the stellar contribution of the majority of the SLRs in a self-consistent manner falls short in terms of stellar contribution of $^{36}$Cl to the ESS. The production of $^{36}$Cl along with $^{7,10}$Be can be explained by local irradiation scenario(s) in the ESS (Leya, Halliday & Wieler 2003; Gounelle et al. 2006; Sahijpal & Gupta 2009).

Finally, in order to understand the influence of the prolonged star formation within a single stellar cluster on the SLRs yields we have performed simulations (Models CI and DI) with distinct star formation histories (Table 1). The star formation within a cluster can linger on for ~3-10 Myr. The SLRs abundance distributions in Models CI and DI are not much different from the Models AI, BI, and EI in which the entire cluster is assumed to have formed within a single timestep of 1 Myr. The proposed hypothesis presented above in the case of Model AI (fig. 3c) can also be applied to any of the *Heter*-models presented in table 1 and figs. 4-7c, 11c, 12c. We have however not marked any thick grey dashed lines unlike those in the fig. 3c in any of the remaining figures for other models.

5. **Conclusion**

We present the galactic chemical evolution model of the steady-state abundance distribution evolution of the SLRs, $^{26}$Al, $^{36}$Cl, $^{41}$Ca, $^{53}$Mn and $^{60}$Fe for the entire Milky Way galaxy over the galactic time scale. This is achieved by developing homogeneous and heterogeneous GCE models with a distinct set of stellar nucleosynthetic yields obtained by different groups. In the former model, we numerically simulate the entire evolution of the galaxy by dividing it into eight annular rings of width 2 kpc each. This model provides a broad average evolutionary behavior of the galaxy in terms of production and evolution of SLRs abundance distribution.

In the heterogeneous GCE models, we simulate the formation and evolution of the successive generations of stellar clusters in an evolving galaxy within small confined pockets having an area of <1 kpc$^2$. The massive star contributions of SLRs are considered in the heterogeneous GCE models towards the evolving inventories of SLRs. We demonstrate the feasibility of the formation of the solar system, along with the inventories of SLRs, within stellar clusters formed during the evolution of the galaxy in terms of sequel episodes of formation and evolution of successive generations of stellar clusters.

Based on our heterogeneous galactic chemical evolution model, we envisage an astrophysical scenario for the stellar origin of most of the SLRs in a self-consistent manner. We hypothesize the possibility that prior to the formation of the stellar cluster-B that contained the sun, another stellar cluster-A was probably formed in its neighborhood perhaps at a distance of several tens of parsec around 25 Myr. ago. The decaying remnants of $^{53}$Mn and $^{60}$Fe from the evolved massive stars from the cluster-A might have contaminated the local ISM associated with the presolar molecular cloud. The Wolf-Rayet winds from a massive star in another stellar cluster-C, at a distance of ~20 parsecs, probably contributed the short-lived nuclides, $^{26}$Al (and $^{41}$Ca) to the presolar molecular cloud. The irradiation production of $^{36}$Cl and $^{7,10}$Be occurred later in the early solar system.

**Acknowledgements:** We are extremely grateful to Dr. R. Trappitsch for several comments and suggestions that led to a substantial improvement of this manuscript. This includes detailed advice regarding recent stellar nucleosynthetic yields by several groups. This work is supported by PLANEX (ISRO) research grant.

**Figure caption (Main text):**

Figure 1. The *Homo*-GCE predictions for the evolutionary trends of SLRs, $^{26}$Al, $^{36}$Cl, $^{41}$Ca, $^{53}$Mn and $^{60}$Fe with Model I (table 1) in the solar annular ring located at 8-10 kpc over the galactic time scale are presented in the form of, a) the ratios of SLRs with respect to stable isotopes, $^{26}$Al/$^{27}$Al, $^{36}$Cl/$^{35}$Cl, $^{41}$Ca/$^{40}$Ca, $^{53}$Mn/$^{55}$Mn and $^{60}$Fe/$^{56}$Fe,  b) the absolute abundance (M$_\odot$ pc$^{-2}$) of $^{26}$Al, $^{36}$Cl, $^{41}$Ca, $^{53}$Mn and $^{60}$Fe. c) The temporal evolution of $^{26}$Al/$^{27}$Al and $^{60}$Fe/$^{56}$Fe across the galaxy for all the annular rings. d-h) The predicted absolute abundance evolution (M$_\odot$ pc$^{-2}$) of the SLRs across the galaxy for all eight annular rings from 2-18 kpc. The assumed solar initial reference values of the SLRs in the early solar system formed around ~9.2 Gyr. are marked in the figure 1a & 1c by solid circles and horizontal bars, respectively.

Figure 2. Same as fig. 1 for the *Homo*-GCE Model IV (table 1).

Figure 3. The deduced, a) star formation (*in the form of stellar associations/clusters*) rates, b) supernovae (SN II, SN Ia, SN Ib/c) rates, c) SLR abundance ratios, and d) SLRs abundances due to massive stars (>11 M$_\odot$) contributions for the *Heter*-GCE Model AI corresponding to a single grid. The formation and evolution of several distinct stellar associations around 9.2 Gyr. are presented in the inset figure. The frequency of the formation and the mass of the stellar associations are determined according to random number generators in the adopted Monte Carlo approach. An average mass of ~3×10$^5$ M$_\odot$ is obtained beyond 9 Gyr for the cluster. $\eta$ and $\chi$ are two scalable parameters that are taken to unity in the figure. However, as discussed in the text these are variable parameters. The assumed solar initial reference values of the SLRs are marked according to table 2 in the fig. 3c. The three thick dashed lines in fig. 3c indicates the approximate match with the initial $^{53}$Mn/$^{55}$Mn and $^{60}$Fe/$^{56}$Fe values in the ESS at the time of the formation of CAIs. This could explain three possible ways for the contribution of the two radionuclides, $^{53}$Mn and $^{60}$Fe, from already evolved stellar clusters to the presolar cloud, with an independent contribution of $^{26}$Al (and $^{41}$Ca) from a Wolf-Rayet star (Sahijpal & Soni 2006; Dwarkadas et al. 2017).

Figure 4. Same as fig. 3 for Model CI of *Heter*-GCE simulation.

Figure 5. Same as fig. 3 for Model DI of *Heter*-GCE simulation.

Figure 6. Same as fig. 3 for Model DIII of *Heter*-GCE simulation.

Figure 7. Same as fig. 3 for Model DIV of *Heter*-GCE simulation.

Figure 8. The predicted evolutionary trends of, a) the metallicity Z, and b) [Fe/H] of the *Heter*-GCE Model BI. The trends are presented for eight randomly selected grids from the solar annular ring over the galactic time scale. The predicted behavior for Z and [Fe/H] shows the maximum possible heterogeneous behavior. The assumed metallicity at the time of the formation of the solar system is assumed to 0.0143 (Asplund et al. 2009).

**Table 1:** The assumptions related to the various simulation parameters for the *Homo*- GCE and *Heter*- GCE models.

| *Homo*-Models | Figure | Stellar nucleosynthetic prescription for the GCE models | | | |
|---|---|---|---|---|---|
| Model I | Fig. 1 | Core-collapse supernova data taken from Woosley & Weaver (1995; WW95); AGB star data (including $^{26}$Al) taken from Karakas (2010; K10); SLR data ($^{41}$Ca, $^{36}$Cl, $^{60}$Fe) taken from http://fruity.oa-teramo.inaf.it/ | | | |
| Model II | Fig. 9 (suppl.) | Core-collapse supernova data taken from WW95; AGB star data taken from K10; SLR data ($^{26}$Al, $^{41}$Ca, $^{36}$Cl, $^{60}$Fe) taken from http://fruity.oa-teramo.inaf.it/ | | | |
| Model III | Fig. 10 (suppl.) | Core-collapse supernova data taken from zero rotational velocity stars from Limongi & Chieffi (2018; LC18); AGB star data (including $^{26}$Al) taken from K10; SLR data ($^{41}$Ca, $^{36}$Cl, $^{60}$Fe) taken from http://fruity.oa-teramo.inaf.it/ | | | |
| Model IV | Fig. 2 | Core-collapse supernova data taken from 300 km s$^{-1}$ rotational velocity stars from LC18; AGB star data (including $^{26}$Al) taken from K10; SLR data ($^{41}$Ca, $^{36}$Cl, $^{60}$Fe) taken from http://fruity.oa-teramo.inaf.it/ | | | |
| *Heter*-Models$^€$ | | Total grids within solar annular ring | Grid size (kpc$^2$) | Time period for the formation of a cluster within a grid | Average mass (M$_\odot$) of stellar cluster formed after ~9 Gyr ($\eta$ = 1) Range is also presented |
| Model AI (WW95+K10) | Fig. 3 | 100 | 1.13 | 50-150 Myr *Entire cluster forms within 1 Myr* | 3.3×10$^5$ Range: 3×10$^4$ - 6×10$^5$ |
| Model BI (WW95+K10) | Fig. 11 (suppl.) | 400 | 0.28 | 25-75 Myr *Entire cluster forms within 1 Myr* | 4.2×10$^4$ Range: 2×10$^3$ - 1×10$^5$ |
| Model CI (WW95+K10) | Fig. 4 | 400 | 0.28 | 50-100 Myr *Entire cluster forms over 3 Myr* | 6.3×10$^4$ Range: 1×10$^4$ - 1×10$^5$ |
| Model DI (WW95+K10) | Fig. 5 | 400 | 0.28 | 50-100 Myr *Entire cluster forms over 10 Myr with a gap of 1 Myr after each 2 Myr* | 8.4×10$^4$ Range: 4×10$^4$ - 1×10$^5$ |
| Model DIII (LC18+K10) 0 km s$^{-1}$ rotational velocity | Fig. 6 | 400 | 0.28 | 50-100 Myr *Entire cluster forms over 10 Myr with a gap of 1 Myr after each 2 Myr* | 8.4×10$^4$ Range: 4×10$^4$ - 1×10$^5$ |
| Model DIV (LC18+K10) 300 km s$^{-1}$ rotational velocity | Fig. 7 | 400 | 0.28 | 50-100 Myr *Entire cluster forms over 10 Myr with a gap of 1 Myr after each 2 Myr* | 8.4×10$^4$ Range: 4×10$^4$ - 1×10$^5$ |
| Model EI (WW95+K10) | Fig. 12 (suppl.) | 800 | 0.14 | 50-110 Myr *Entire cluster forms within 1 Myr* | 3.3×10$^4$ Range: 6×10$^3$ - 7×10$^4$ |

$€$ The SLR stellar yields of the core-collapse supernovae were only considered for *Heter*-Models

**Table 2:** The predicted ratios of the short-lived radionuclides (along with their mean-life) with respect to their stable nuclides for the various *Homo*-GCE models (table 1) at the time of the formation of the solar system ~9.2 Gyr in 8-10 kpc annular ring are compared with the assumed ESS reference values.

| | Model I | Model II | Model III | Model IV | Assume initial value in the early solar system | Reference |
|---|---|---|---|---|---|---|
| $^{26}$Al/$^{27}$Al ($\tau$ ~1.05 Myr) | $4.4\times10^{-6}$ *$2.6\times10^{-6\,\$}$* | $2.7\times10^{-6}$ *$2.1\times10^{-6\,\$}$* | $4.6\times10^{-6}$ *$2.3\times10^{-6\,\$}$* | $6.3\times10^{-6}$ *$3.7\times10^{-6\,\$}$* | $(5.23\pm0.13)\times10^{-5}$ | Jacobsen et al. (2008) |
| $^{36}$Cl/$^{35}$Cl ($\tau$ ~0.43 Myr) | $2.8\times10^{-6}$ | $2.8\times10^{-6}$ | $2.2\times10^{-6}$ | $3.7\times10^{-6}$ | $(2.44\pm0.65)\times10^{-5}$ | Tang et al. (2017) |
| $^{41}$Ca/$^{40}$Ca ($\tau$ ~0.15 Myr) | $3.9\times10^{-7}$ | $3.9\times10^{-7}$ | $2.0\times10^{-7}$ | $5.3\times10^{-7}$ | $(4.6\pm1.9)\times10^{-9}$ | Liu (2017) |
| $^{53}$Mn/$^{55}$Mn ($\tau$ ~5.34 Myr) | $6.7\times10^{-5}$ | $6.8\times10^{-5}$ | $7.7\times10^{-5}$ | $8.9\times10^{-5}$ | $(7\pm1)\times10^{-6}$ | Tissot et al. (2017) |
| $^{60}$Fe/$^{56}$Fe ($\tau$ ~3.75 Myr) | $1.2\times10^{-7}$ *$9.2\times10^{-8\,\$}$* | $1.2\times10^{-7}$ *$9.2\times10^{-8\,\$}$* | $7.9\times10^{-8}$ *$6.4\times10^{-8\,\$}$* | $2.1\times10^{-7}$ *$1.7\times10^{-7\,\$}$* | $(1.0\pm0.27)\times10^{-8}$ | Tang & Dauphas (2015) |

$ Present day GCE estimates based on fig. 1a, 2a, 9a & 10a

**References:**

Amelin Y., 2008, GCA, 72, 221

Amelin Y., Kaltenbach A., Iizuka T., Stirling C.H., Ireland, T.R., Petaev, M., Jacobsen, S.B., 2010, Earth Planet. Sci. Lett., 300, 343

Asplund M., Grevesse N., Sauval A. J., Scott P., 2009, ARA&A, 47, 481

Bhatia G. K., Sahijpal S., 2016, Meteorit. Planet. Sci., 51, 138

Bizzarro M., Baker J.A., Haack H., Nature, 2005, 435, 1280

Bojazi M.J., Meyer B.S., 2017, in 80[th] Annual Meeting of the Meteorit. Soc., Vol. 1987

Boss A.P., 1995, ApJ, 439, 224

Boss A.P., 2017, ApJ, 844, 113

Bouvier A., Wadhwa M., 2010, Nat. Geosci., 3, 637

Brennecka G. A., Wadhwa M., 2012, Proc. Nat. Acad. Sci. USA, 109, 9299

Cameron A. G. W., Höflich P., Myers P. C., Clayton D. D., 1995, ApJ, 447, 53

Chiappini C., Matteucci F., Gratton R., 1997, ApJ, 477, 765

Chiappini C., Matteucci F., Romano D., 2001, ApJ, 554, 1044

Clark P. C., Glover S. C. O., Klessen R. S., Bonnell I. A., 2012, MNRAS, 424, 2599

Clayton D. D., Hartmann D. H., Leising M. D., 1993, ApJ, 415, L25

Connelly J.N., Bizzarro M., Krot A.N., Nordlund Å., Wielandt D., Ivanova, M.A., 2012, Science, 338, 651

Cristallo S., Straniero O., Piersanti L., Gobrecht D., 2015 ApJSS, 219, 40

Dauphas N., Chaussidon M., 2011, Ann. Rev. Earth Planet. Sci., 39, 351

de Avillez M. A., Mac Low M.-M., ApJ, 2002, 581, 1047

Diehl R. 2016, J. Phys. Conf. Ser., 665, 012011

Diehl R. et al., 2006a, Nature, 439, 45

Diehl R. et al., 2006b, A&A, 449, 1025

Diehl R. et al. 2010, A&A, 522, A51

Dwarkadas V.V., Dauphas N., Meyer B., Boyajian P., Bojazi M., 2017, ApJ, 851, 147

Fall S. M., Chandar R., 2012, ApJ, 752, 96

Feigelson E.D., Townsley L.K., 2008, ApJ, 673, 354

Fujimoto Y., Krumholz M. R., Tachibana S., 2018, MNRAS, 480, 4025

Gaidos E., Krot A. N., Williams J. P., Raymond S. N., 2009, ApJ, 696, 1854

Galy A., Hutcheon I. D., Grossman L., 2004, in 35th Lunar & Planet. Sci. Conf., id. 1790

Glavin D. P., Kubny A., Jagoutz E., Lugmair G. W., 2004, Meteorit. Planet. Sci. 39, 693

Göpel C., Manhes G., Allegre C. J., 1994, Earth Planet Sci. Lett., 121, 153

Gounelle M., Shu F.H., Shang H., Glassgold A.E., Rehm K.E., Lee T. 2006. ApJ, 640, 1163

Gounelle M., Meibom A., Hennebelle P., Inutsuka S., 2009, ApJ, 694, L1

Grimm R.E., McSween H.Y., 1993, Science, 259, 653

Grisoni V., Spitoni E., Matteucci F., 2018, MNRAS, 481, 2570

Hester J.J., Healy K.R., Desch S.J., 2004, in American Astron. Soc. Meeting 205, Bull. Am. Astron. Soc., 36, 1516

Howard C. S., Pudritz R. E., Harris W. E., 2014, MNRAS, 438, 1305

Huss G. R., Meyer B. S., Srinivasan G., Goswami J. N., Sahijpal S., 2009, GCA, 73, 4922

Ito M., Nagasawa H., Yurimoto H., 2006, Meteorit. Planet. Sci., 41, 1871

Iwamoto K., Brachwitz F., Nomoto K.I., Kishimoto N., Umeda H., Hix W.R., Thielemann F.K., 1999, ApJS, 125, 439

Jacobsen B., Yin Q.Z., Moynier F., Amelin Y., Krot A.N., Nagashima K., Hutcheon I.D., Palme H., 2008, Earth Planet. Sci. Lett., 272, 353

Jones S.W. et al., 2019, MNRAS, 485, 4287

Karakas A. I, Lattanzio J.C., 2003, PASA, 20, 279

Karakas A. I., 2010, MNRAS, 403, 1413

Kastner J.H., Myers P.C., 1994, ApJ, 421, 605

Kaur T., Sahijpal S., 2019, in 50th Lunar & Planet. Sci. Conf., id. 1932

Kobayashi C., Umeda H., Nomoto K., Tominaga N., Ohkubo T., 2006, ApJ, 653, 1145

Krot A. N., Chaussidon M., Yurimoto H., Sakamoto N., Nagashima K., Hutcheon I. D., MacPherson, G. J., 2008, GCA, 72, 2534

Krot A.N., Makide, K. Nagashima, K., Huss G.R., Ogliore R.C., Ciesla F.J., Yang L., Hellebrand E., Gaidos E., 2012, Meteorit. Planet. Sci., 47, 1948

Kroupa P., 2008, in de Koter A., Smith L., Waters R. San Francisco, eds, ASP Conf. Ser. Vol. 388. Mass loss from stars and the evolution of stellar clusters Astron. Soc. Pac., San Francisco, p. 271

Lada C. J., Lada E. A., 2003, ARA&A, 41, 57

Lada C. J., 2010, Phil. Trans. R. Soc. A, 368, 713

Lada C. J., Lombardi M., Alves J. F., 2010, ApJ, 724, 687

Lee T., Papanastassiou D.A., Wasserburg G.J., 1976, Geophys. Res. Lett., 3, 41

Lee T., Papanastassiou D.A., Wasserburg G.J., 1977, ApJ, 211, L107

Leya I., Halliday A. N., Wieler R., 2003, ApJ, 594, 605

Limongi M., & Chieffi A., 2012, ApJ, 199, 9

Limongi M., Chieffi A., 2018, ApJSS, 237, 13

Liu M-.C., 2017, GCA, 201, 123

Liu M. C., Chaussidon M., Srinivasan G., McKeegan K. D., 2012, ApJ, 761, 137

Lugaro M., Ott U., Keresztüri Á., 2018, Prog. Part. Nucl. Phys., 102, 1

Lugmair G. W., Galer S. J. G., 1992, GCA, 56, 1673

Lugmair G. W., Shukolyukov A., 1998, GCA, 62, 2863

MacPherson G.J., Davis A.M., Zinner E.K., 1995, Meteoritics, 30, 365

Martin P., Knödlseder J., Diehl, R., Meynet G., 2009, A&A, 506, 703

Matteucci F., François P., 1989, MNRAS, 239, 885

Matteucci F., 2003, The Chemical Evolution of the Galaxy. Kluwer Academic Publishers, Dordrecht, ZH

Matteucci F., Spitoni E., Recchi S., Valiante R., 2009, A&A, 501, 531

Merk R., Breuer D., Spohn T., 2002, Icarus, 159, 183

Micali A., Matteucci F., Romano D., 2013, MNRAS, 436, 1648

Minchev I., Famaey B., Quillen A.C., Dehnen W., Martig M., Siebert A., 2012, A&A, 548, 127

Minchev I., Martig M., Streich D., Scannapieco C., de Jong R. S., Steinmetz M., 2015, ApJ, 804, L9

Mishra R. K., Chaussidon M. 2014, Earth Planet. Sci. let., 398, 90

Mishra R.K., Goswami J.N., 2014, GCA 132, 440

Mishra R.K., Marhas K.K., Sameer, 2016, Earth Planet. Sci. let., 436, 71

Mostefaoui S., Lugmair G.W., Hoppe P., 2005, ApJ, 625, 271.

Murray N., 2011, ApJ, 729, 133

Neumann W., Breuer D., Spohn T., 2014, Earth Planet. Sci. Lett., 395, 267

Ouellette N., Desch S. J., Hester J. J., 2007, ApJ, 662, 1268

Pagel B.E.J., 1997, Nucleosynthesis and Chemical Evolution of Galaxies. Cambridge Univ.

Press, Cambridge, UK

Pfalzner S. et al., 2015, Phys. Scr., 90, 068001

Prantzos, N., Diehl, R., 1996, Phys. Rep., 267, 1

Roy J.-R., Kunth D., 1994, A&A, 294, 432

Russell S. S., Gounelle M., Hutchison R., 2001, Philos. Trans. R. Soc. London. Ser. A Math. Phys. Eng. Sci., 359, 1991

Russell S.S., Srinivasan G., Huss G.R., Wasserburg G.J., MacPherson, G.J., 1996, Science, 273, 757

Sahijpal S., 2013, JA&A, 34, 297

Sahijpal S., 2014a, JA&A, 35, 121

Sahijpal S. 2014b, RA&A, 14, 693

Sahijpal S., Gupta G., 2009, Meteorit. Planet. Sci., 44, 879

Sahijpal S., Gupta G., 2013, Meteorit. Planet. Sci., 48, 1007

Sahijpal S., Kaur T., 2018, MNRAS, 481, 5350

Sahijpal S., Goswami J. N., Davis A. M., 2000, GCA, 64, 1989

Sahijpal, S., Soni, P., 2006, Meteorit. Planet. Sci., 41, 953

Sahijpal S., Soni P., Gupta G., 2007, Meteorit. Planet. Sci., 42,1529

Scalo J. M., 1998, in Gilmore G., Howell D., eds, ASP Conf. Ser. Vol. 142, The Stellar Initial Mass Function: 38th Herstmonceux Conference. Astron. Soc. Pac., San Francisco, p. 201

Smith N., 2006, MNRAS, 367, 763

Smith N., Brooks K.J., 2007, MNRAS, 379, 1279

Tang H., Dauphas N., 2012, Earth Planet. Sci. Lett., 359, 248

Tang H., Dauphas N., 2015, ApJ, 802, 22

Tang H., Liu M.C., McKeegan K.D., Tissot F.L., Dauphas N., 2017, GCA, 207,1

Tasker E. J., Tan J. C., 2009, ApJ, 700, 358

Telus M., Huss G. R., Ogliore R. C., Nagashima K., Tachibana S. 2012, Meteoritics & Planet. Sci., 47, 2013

Telus M., Huss G.R., Ogliore R.C., Nagashima K., Howard D.L., Newville M.G., Tomkins A.G., 2016, GCA, 178, 87

Telus M., Huss, G. R., Nagashima, K., Ogliore, R. C., & Tachibana, S. 2018, GCA, 221, 342

Thrane K., Bizzarro M., Baker J.A., 2006, ApJ, 646, L159

Thies I., Kroupa P., Theis C., 2005, MNRAS, 364, 961

Tissot F.L., Dauphas N., Grove, T.L., 2017, GCA, 213, 593

Trappitsch R., et al., 2018, ApJ, 857, L15

Trujillo C. A., Sheppard S. S., 2014, Nature, 507, 471

Urey, H. C. 1955, Proc. Natl. Acad. Sci., 41, 127

Wang W. et al., 2007, A&A, 469, 1005

Wasserburg G.J., Busso M., Gallino R., Nollett K.M., 2006, Nuclear Physics A, 777, 5

Weidner C., Kroupa P., 2006, MNRAS, 365, 1333

Woosley S. E., Weaver T. A., 1995, ApJS, 101, 181

Young E.D., 2014, Earth Planet. Sci. let. 392, 16

Young, E.D., 2016, ApJ, 826, 129

Young E.D., Simon J.I., Galy A., Russell, S.S., Tonui E., Lovera O., 2005, Science, 308, 223

**Figures (Main Text)**

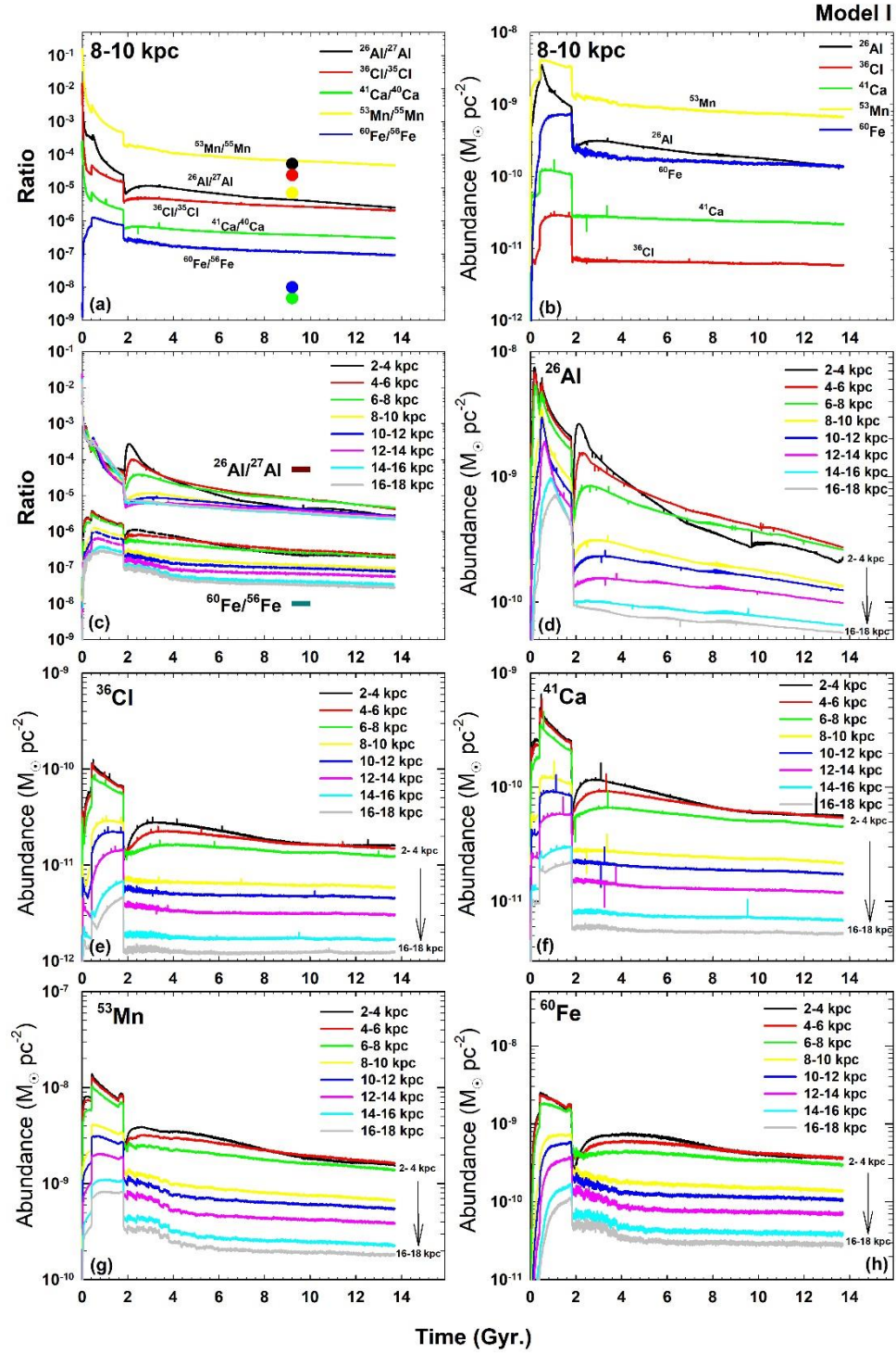

**Fig. 1**

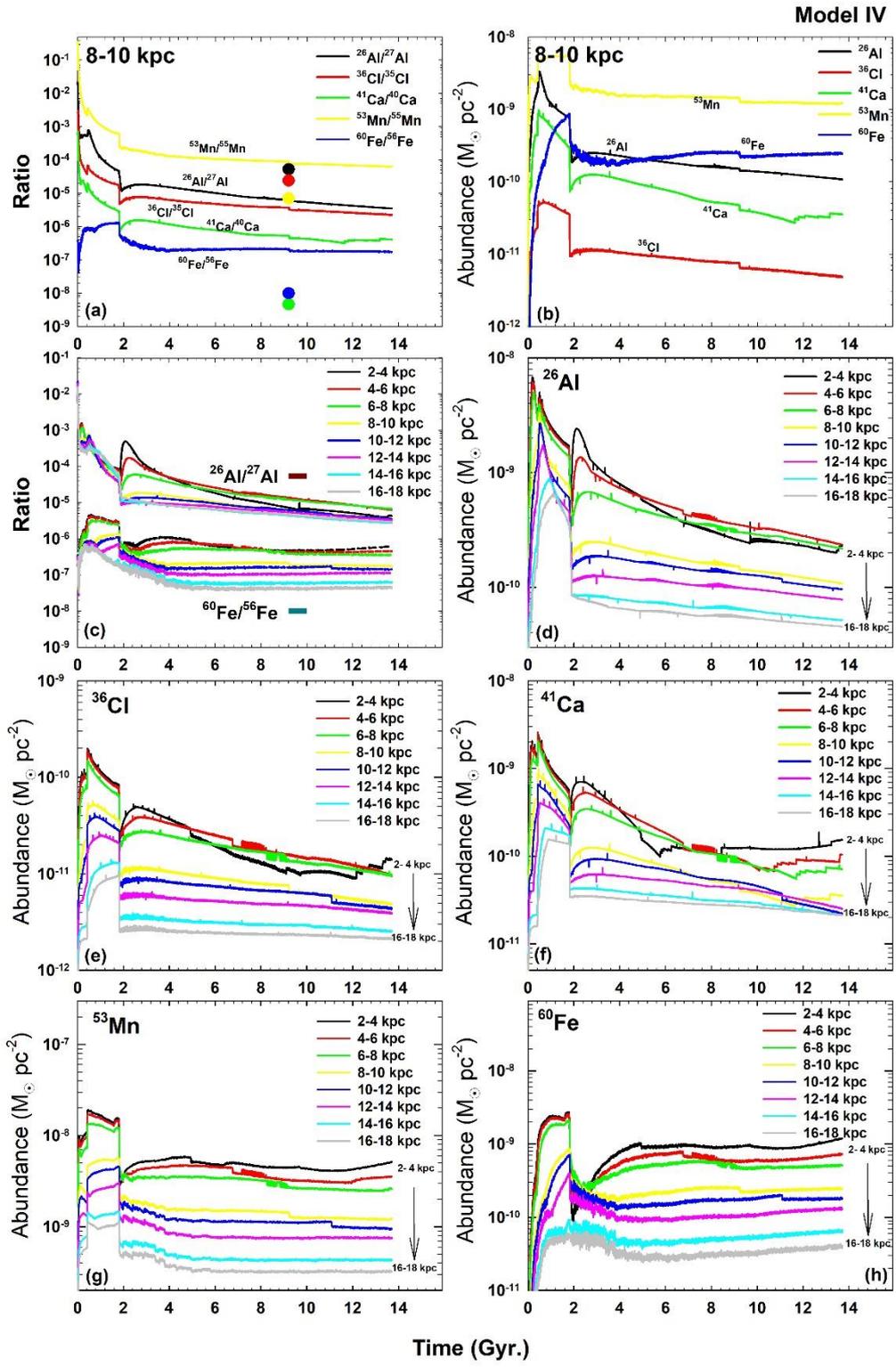

Fig. 2

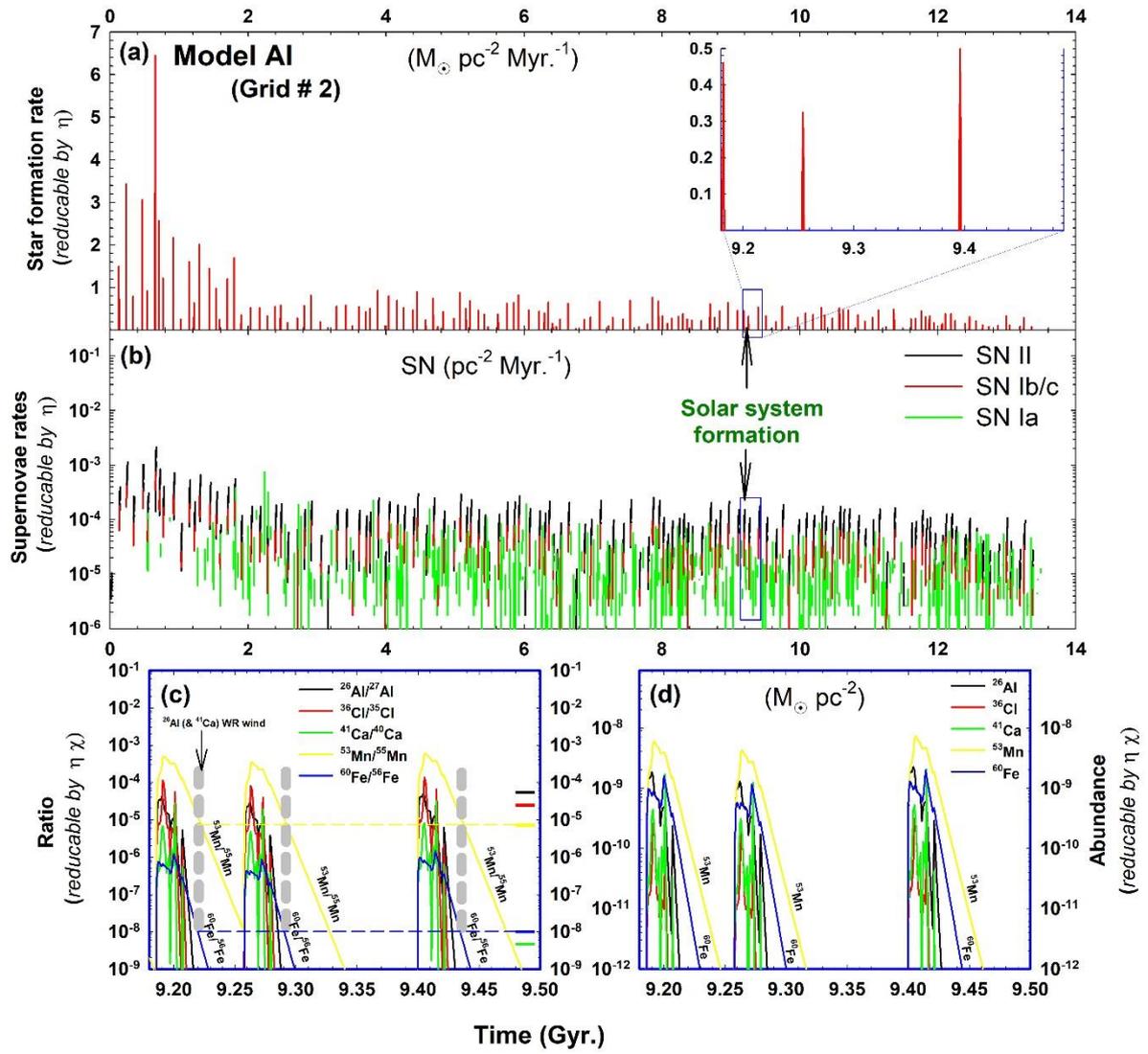

**Fig. 3**

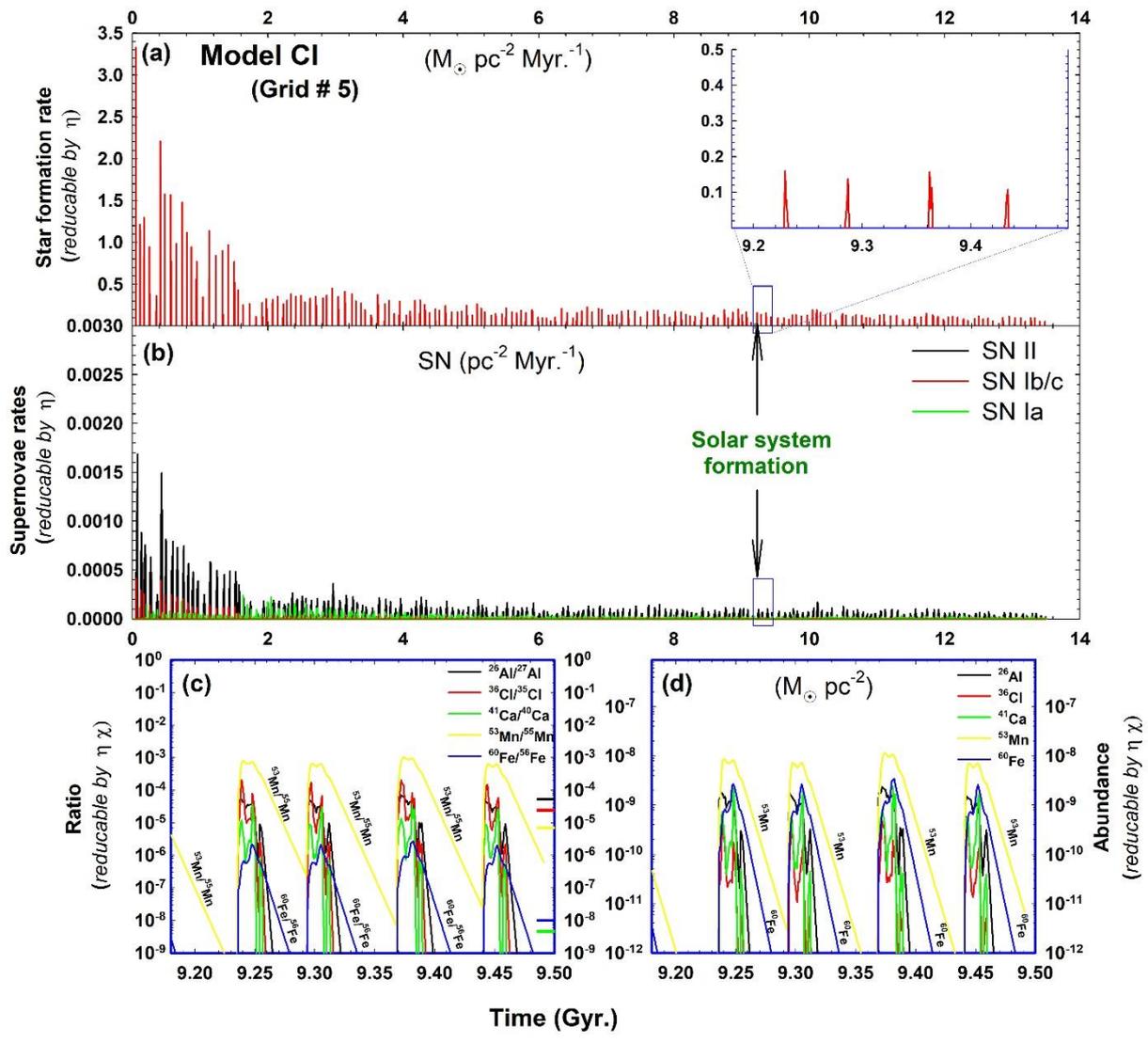

**Fig. 4**

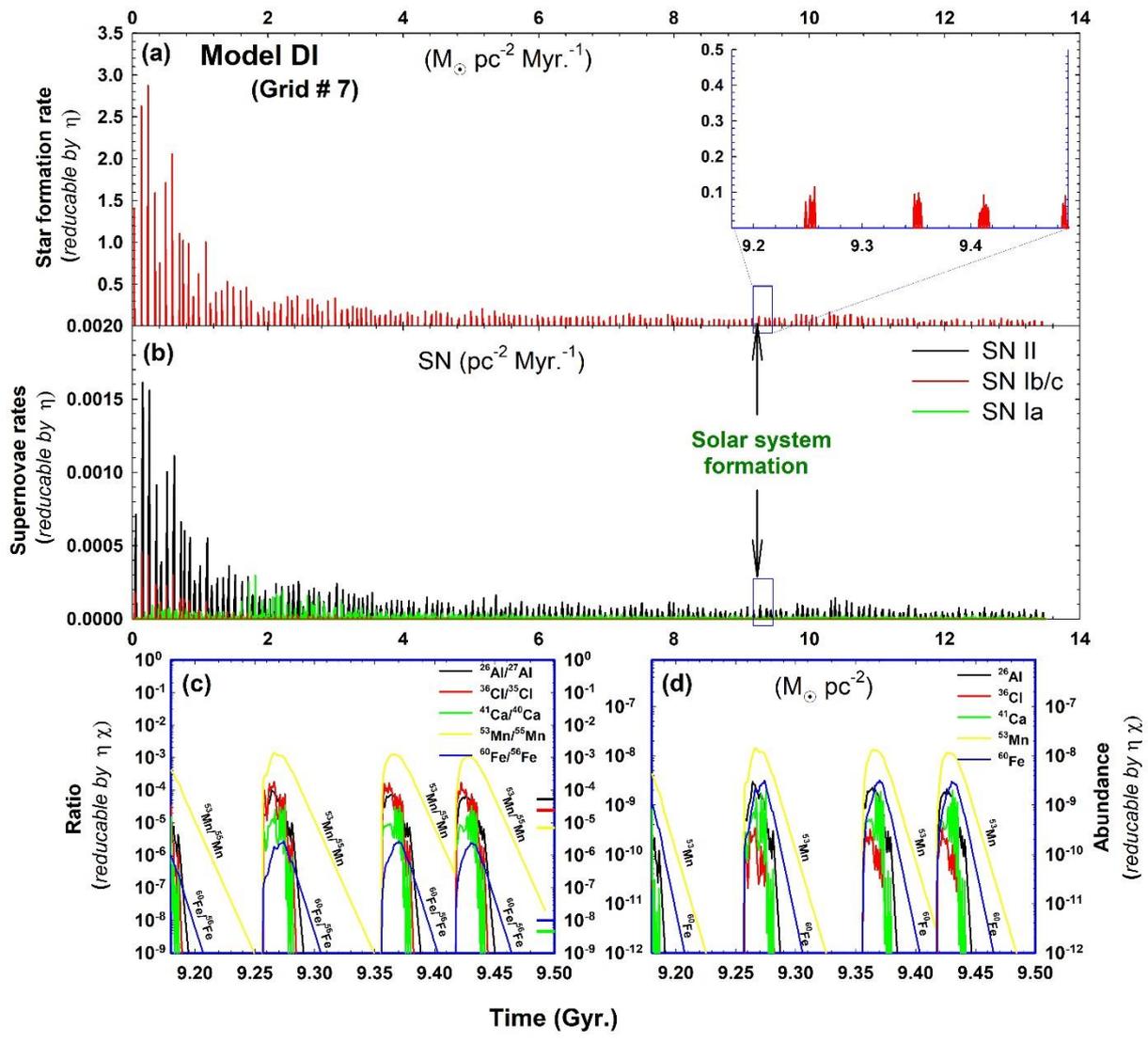

**Fig. 5**

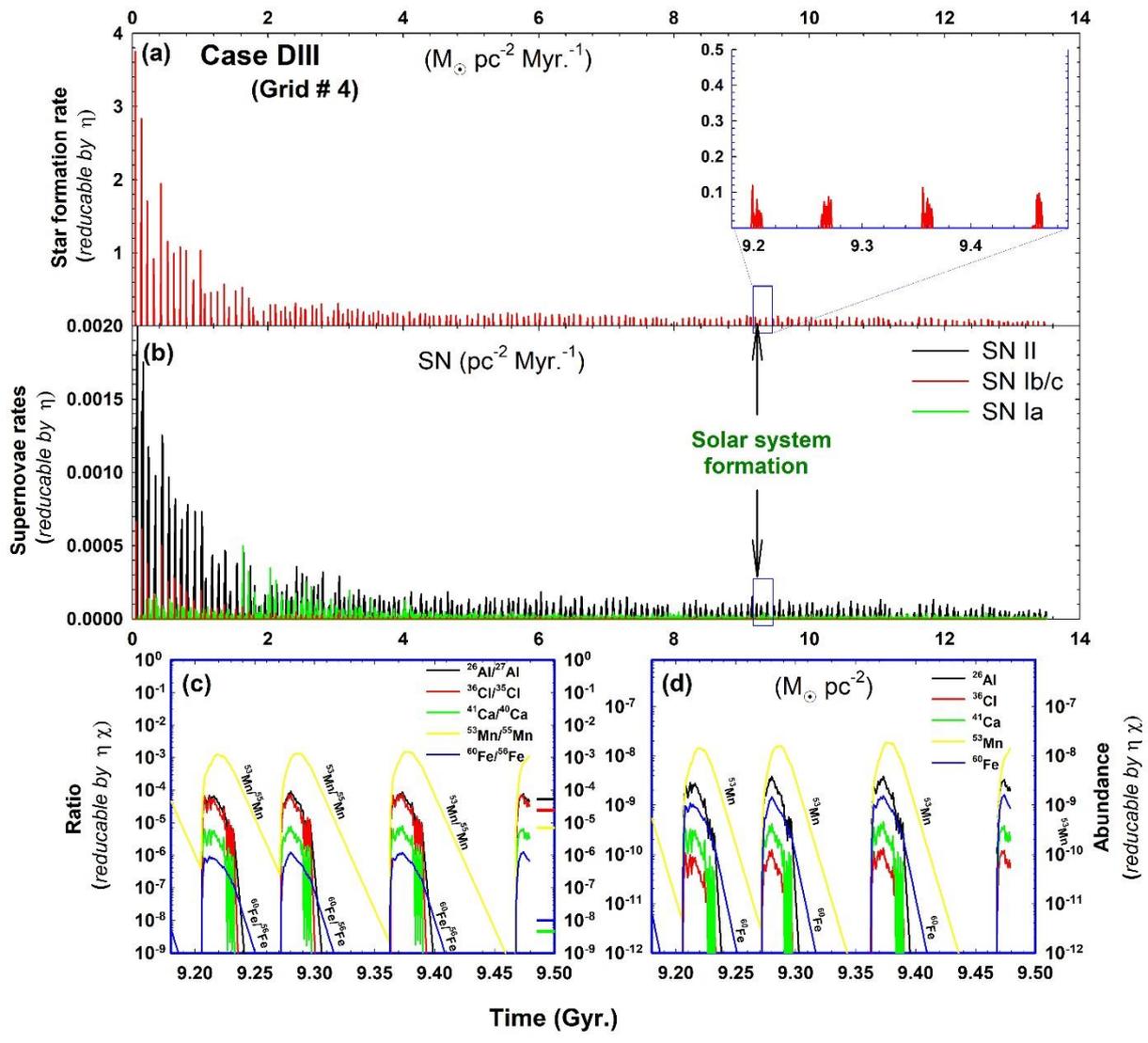

**Fig. 6**

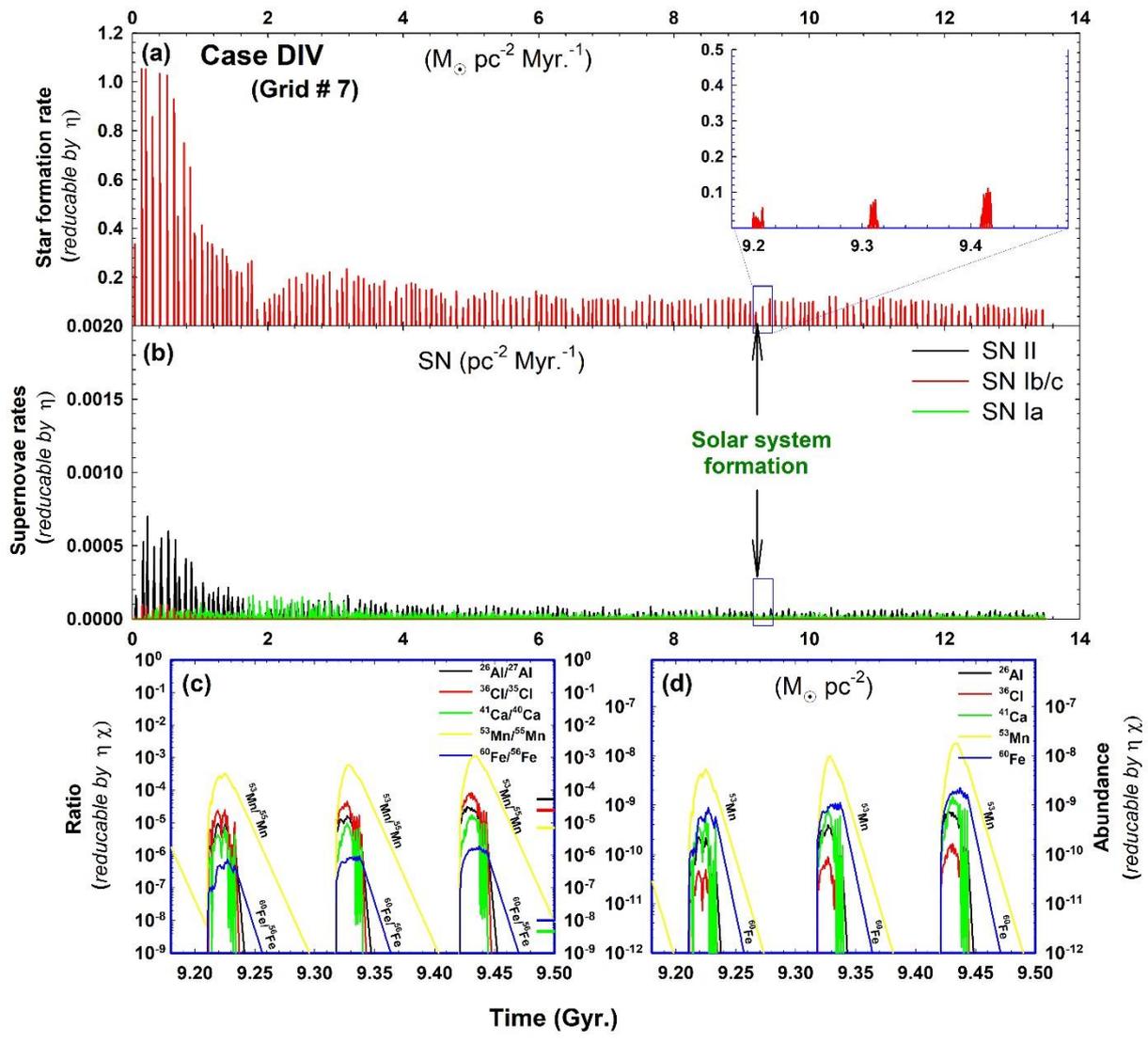

**Fig. 7**

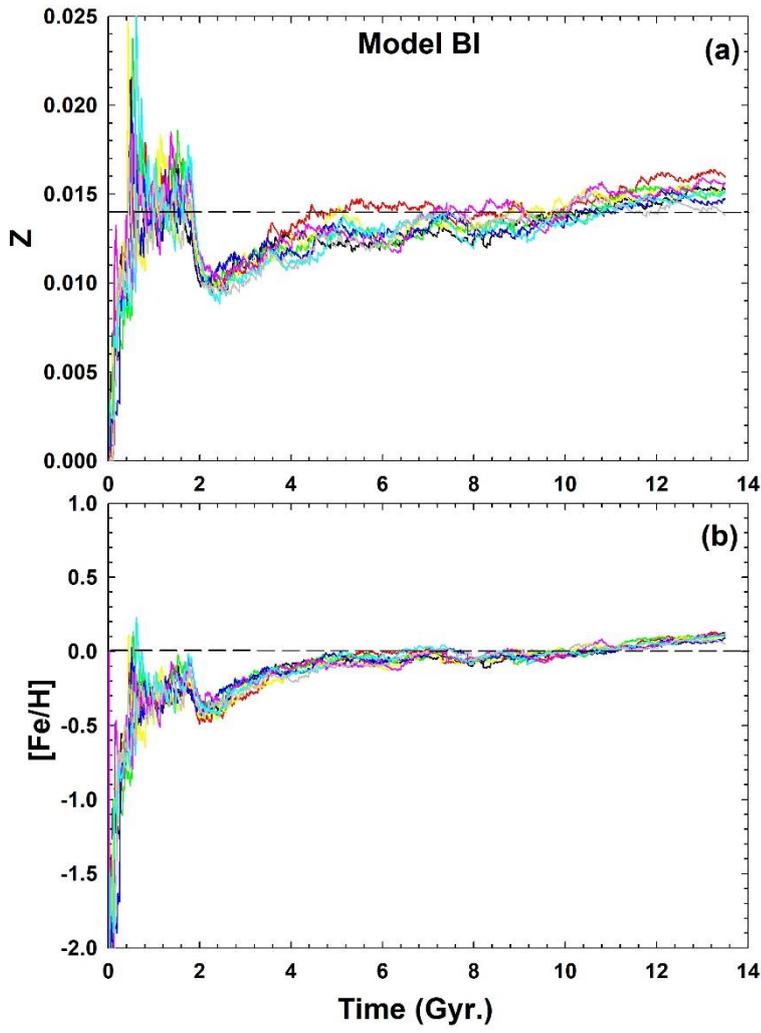

**Fig. 8**

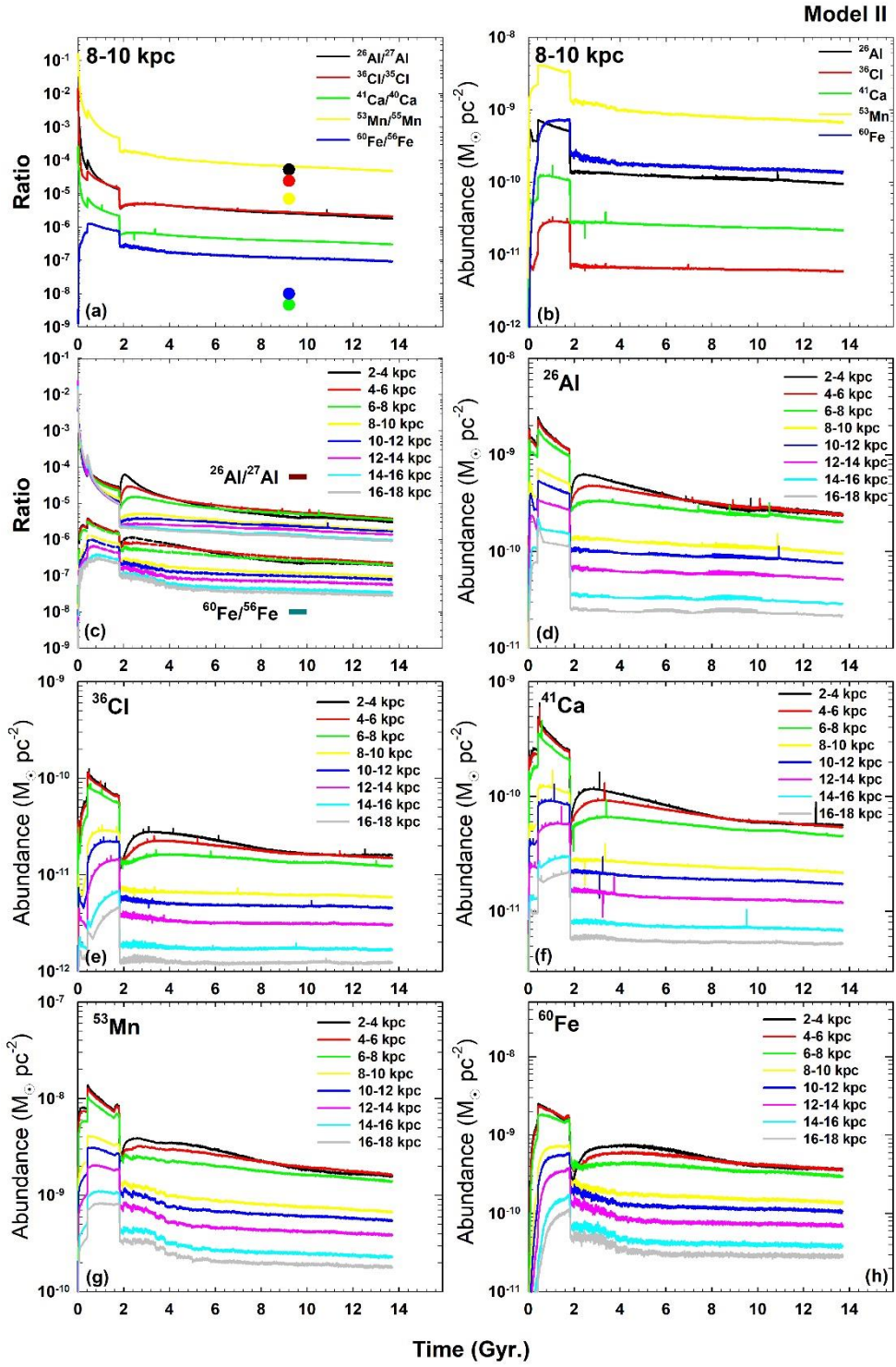

**Figure 9.** Same as fig. 1 for Model II of *Homo*-GCE simulation.

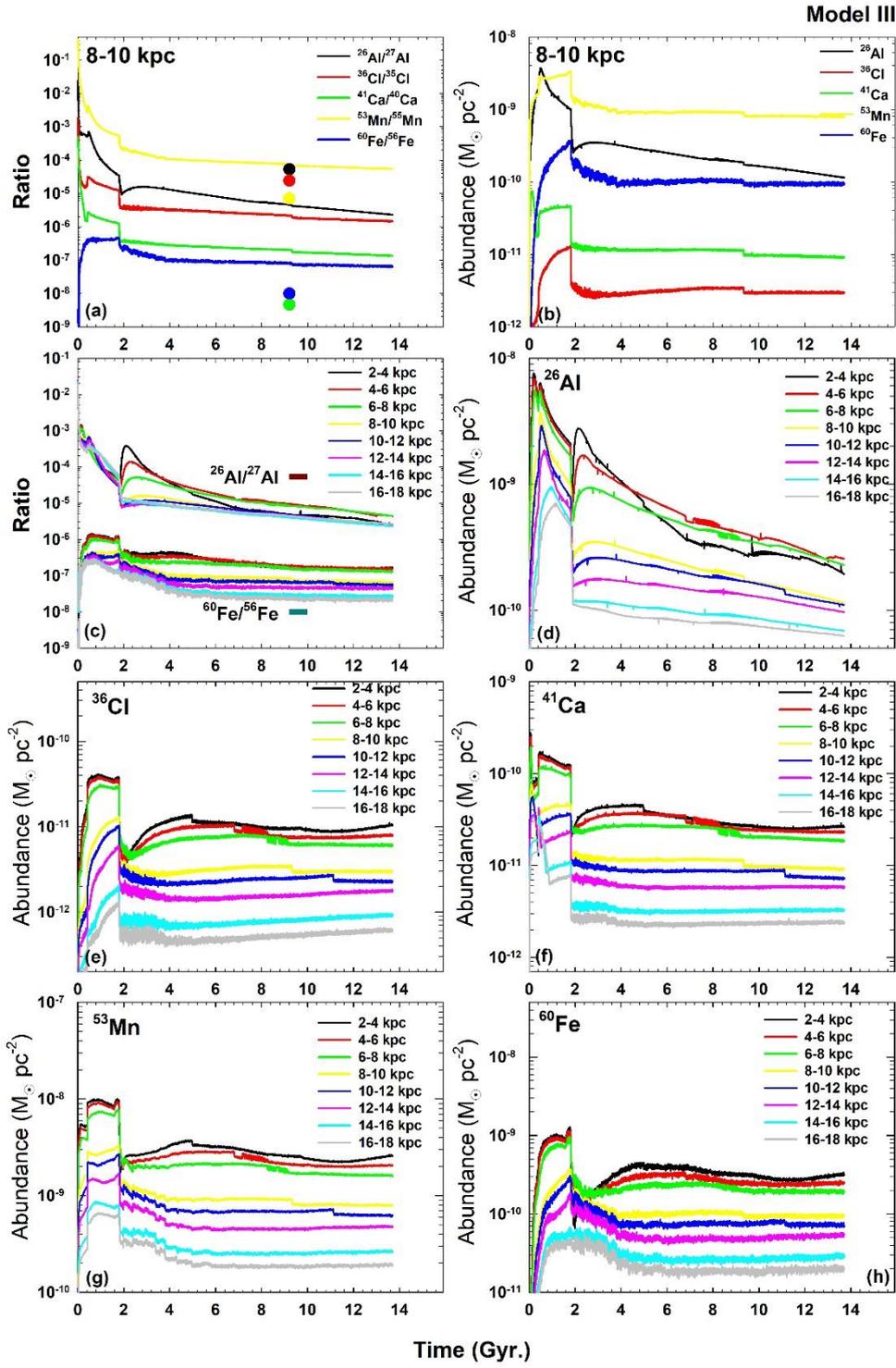

**Figure 10.** Same as fig. 1 for Model III of *Homo*-GCE simulation.

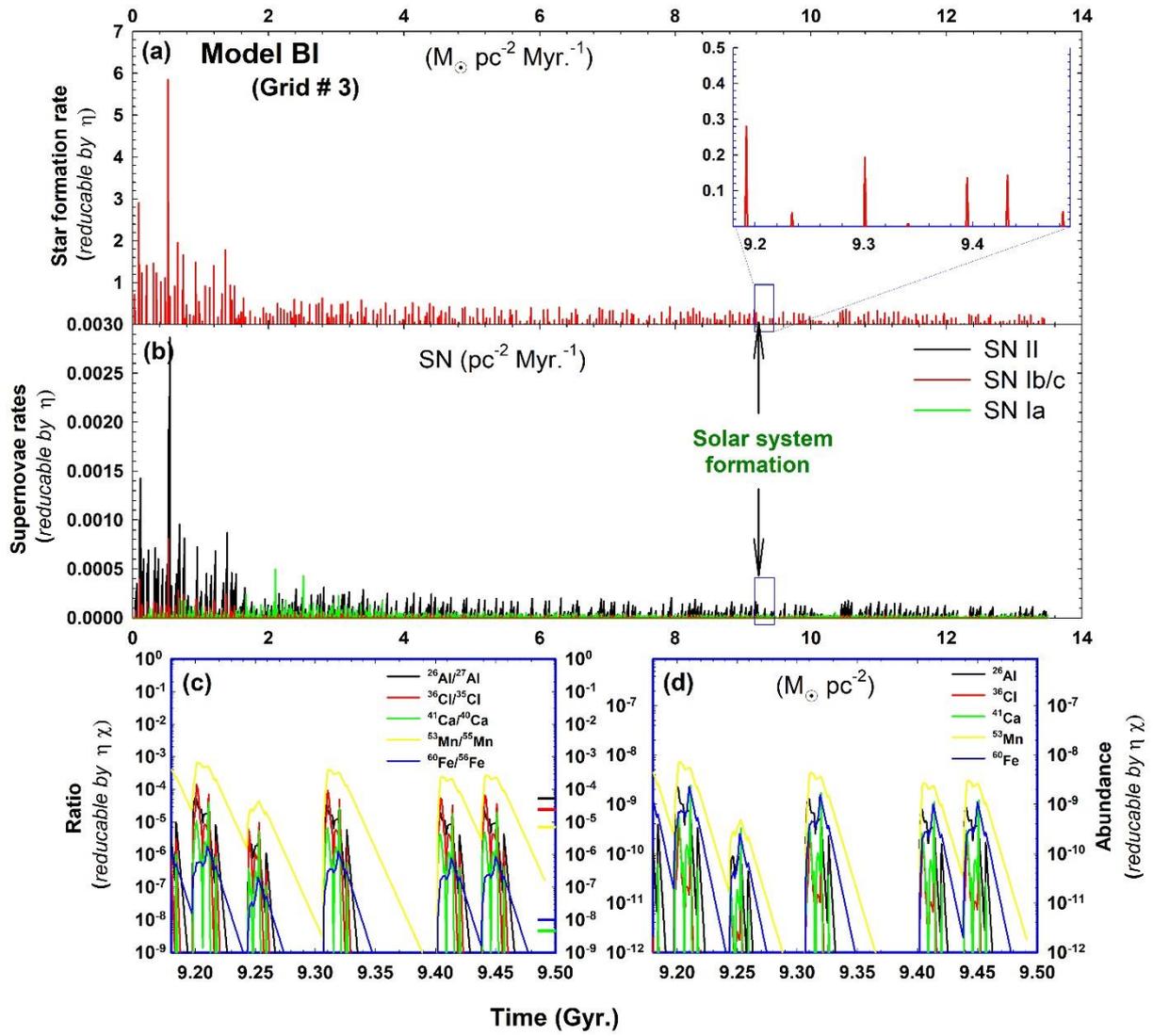

**Figure 11**. Same as fig. 3 for Model BI of *Heter*-GCE simulation.

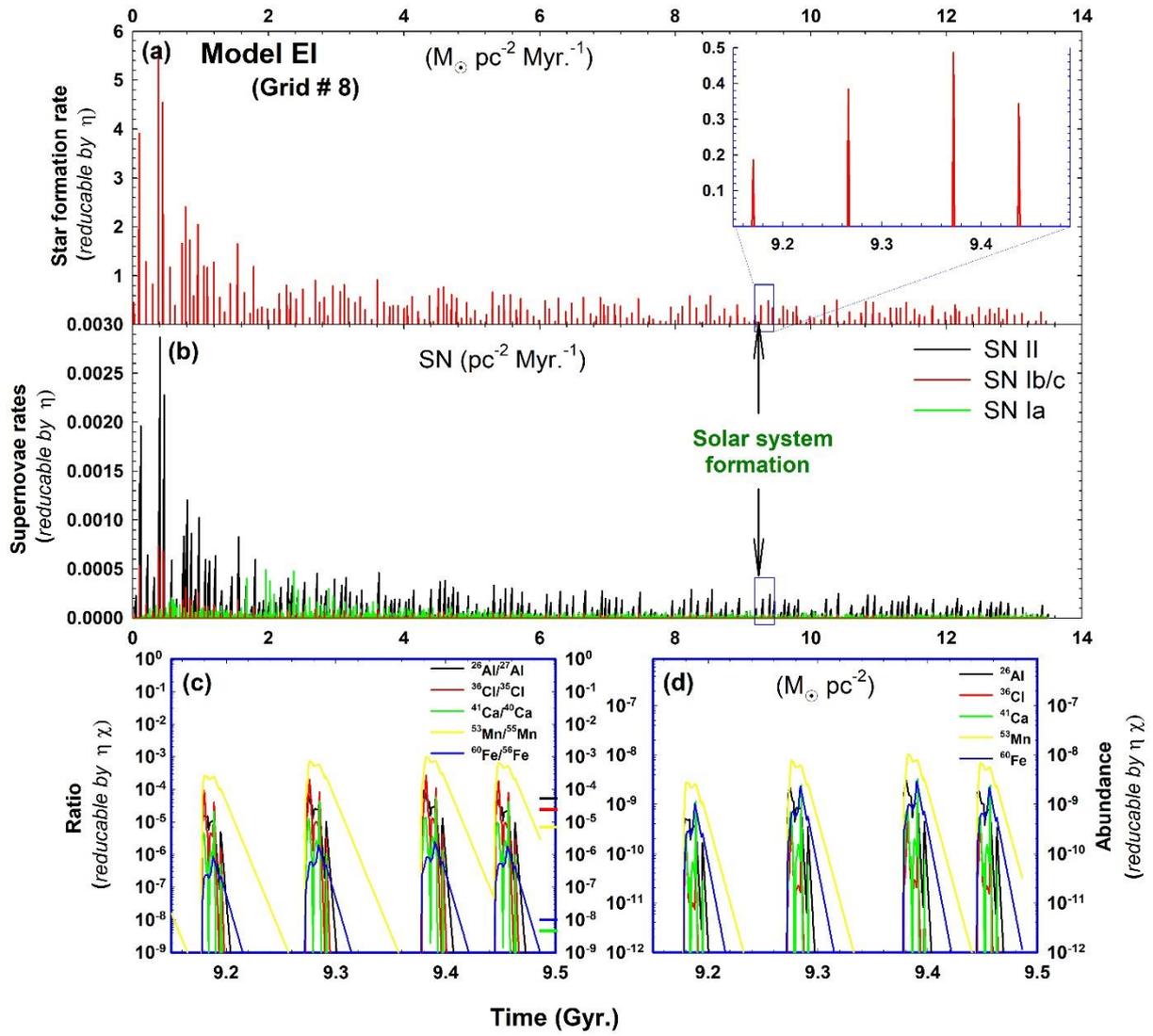

**Figure 12**. Same as fig. 3 for Model EI of *Heter*-GCE simulation.